\documentclass[10pt,times,letter]{article}

\usepackage{amsmath, amsfonts,amsthm,amssymb,bm}
\usepackage{dsfont}

\usepackage[english]{babel}
\DeclareMathAlphabet\mathbfcal{OMS}{cmsy}{b}{n}

\usepackage{latexsym,amssymb,amsfonts,graphicx,bbm}
\usepackage{amsmath,dsfont}
\usepackage{verbatim}
\usepackage{mathrsfs}
\usepackage{bm}
\usepackage{color}
\usepackage{epsfig}

\usepackage{url}
\usepackage{bm}
\usepackage{natbib}

\usepackage[colorlinks,linkcolor=blue,citecolor=blue,anchorcolor=blue]{hyperref}

\newcommand{\Qx}{ \mathbb{Q} }
\newcommand{\N}{ \mathds{N} }
\newcommand{\Ex}{ \mathbb{E} }
\newcommand{\Hx}{\mathbb{H}}

\newcommand{\A}{{\cal A}}

\newcommand{\Ttau}{[0,T\wedge{\tau}_{N+1}]}
\newcommand{\tauN}{\tau_{N+1}}

\newcommand{\Gx}{\mathbb{G}}
\newcommand{\Fx}{\mathbb{F} }

\newcommand{\D}{d}

\newcommand{\F}{\mathcal{F}}
\newcommand{\G}{\mathcal{G}}
\newcommand{\R}{\mathds{R}}

\newtheorem{theorem}{Theorem}[section]
\newtheorem{definition}{Definition}[section]

\newtheorem{prop}[theorem]{Proposition}
\newtheorem{remark}[theorem]{Remark}
\newtheorem{lemma}[theorem]{Lemma}

\newtheorem{example}[theorem]{Example}
\allowdisplaybreaks[4]

\usepackage[normalem]{ulem}

\title
{
Risk-Minimizing Hedging of Counterparty Risk
}

\author{
Lijun Bo \thanks{Email: lijunbo@ustc.edu.cn, School of Mathematics and Statistics, Xidian University, Xi¡¯an 710071, China, and School of Mathematical Sciences, University of Science and Technology of China, Hefei, Anhui
Province, 230026, China.} \and
Agostino Capponi \thanks{Email: ac3827@columbia.edu, Department of Industrial Engineering and Operations Research, Columbia University, New York, 10027, NY, USA} \and
Claudia Ceci \thanks{Email: c.ceci@unich.it, Department of Economics, University "G. d'Annunzio" of Chieti-Pescara, Pescara 65127, Italy.}}

\begin{document}

\maketitle

\begin{abstract}
We study dynamic hedging of counterparty risk for a portfolio of credit derivatives. Our empirically driven credit model consists of interacting default intensities which ramp up and then decay after the occurrence of credit events. {Using the Galtchouk-Kunita-Watanabe decomposition of the counterparty risk price payment stream, we recover} 
a closed-form representation for the risk minimizing strategy in terms of classical solutions to nonlinear recursive systems of Cauchy problems. {We discuss applications of our framework to the most prominent class of credit derivatives, including credit swap and risky bond portfolios, as well as first-to-default claims.}

\end{abstract}

\vspace{0.3 cm}

{\noindent{\textbf{AMS 2000 subject classifications}: 60J25, 60J75, 60H30, 91B28.}}

\vspace{0.3 cm}

\noindent{\textbf{Keywords and phrases}:}\quad {risk minimization; Galtchouk-Kunita-Watanabe decomposition; nonlinear recursive system of Cauchy problems; counterparty risk}.

\section{Introduction}
We study dynamic hedging of counterparty risk for portfolio credit derivatives. The vast majority of literature has focused on the valuation of counterparty risk, abbreviated with CVA throughout this paper; see also \cite{capmig} for a survey. Despite the importance of dynamic hedging of counterparty risk across policy makers and the financial industry, 
the literature on the subject is still not as well developed.\footnote{\cite{cannlessons} argues that the high market volatility experienced during the global financial crisis created challenges for the dynamic hedge of CVA.}
A larger body of literature has investigated dynamic hedging of defaultable claims using mean variance strategies, but without accounting for counterparty risk. \cite{BJR04a} and \cite{BJR04b} introduce a framework for hedging risks in incomplete markets, building on the classical Markowitz mean-variance portfolio selection framework. They analyze quadratic hedging methods and consider strategies adapted to the default-free market information as well as to the enlarged filtration inclusive of default events. \cite{BielAAP} consider a reduced form framework driven by a Brownian motion, and show that perfect hedging can be achieved by continuously trading rolling credit default swap (CDS) contracts. \cite{Freyback} analyze hedging of synthetic CDO tranches under a dynamic credit risk model with incomplete information, allowing for default contagion and spread risk. As in our paper, they use the risk-minimization approach, and choose single name credit swaps as their dynamic trading instruments. 

We study hedging, in the risk-minimization sense, of counterparty risk associated with portfolio credit derivatives traded between a default-free investor and a defaultable counterparty. {Risk-minimization is a quadratic hedging method, {usually applied for derivatives hedging in incomplete financial markets}, which maintains the replicability constraint but relaxes the self-financing condition. Precisely, the risk-minimizing hedging strategy 
{is self-financing on average (mean self-financing)} and minimizes the associated risk measured by the conditional expected value of squared future {hedging} costs. It is strictly connected to the Galtchouk-Kunita-Watanabe (GWK) decomposition of the claim {with respect to risky assets used as hedging instruments.}  The general framework has been introduced by {\cite{FSo}} in the martingale case, and then generalized in \cite{Msfirst} to the semimartingale case. We refer to \cite{guided} for a survey. The methodology has been subsequently extended to include payment streams in \cite{Schweizer08}.

We propose a general model of direct default contagion, which accounts for the impact of past defaults on the default intensity of surviving firms. Our model can be specialized to capture the main sources of default correlation identified by empirical research. For instance, \cite{GieseckeClu} document the time decay effect of default contagion via a statistical analysis based on historical corporate default data. By choosing a linear specification for the default intensity function, after a ramp-up for the instantaneous impact of a default, the default intensities of surviving firms would, over time, mean revert to their long run averages.}

We consider the counterparty risk hedging of a portfolio of defaultable claims of generic type, including classes of credit derivatives routinely used by risk management divisions such as CDSs portfolios, risky bonds portfolios and first-to-default claims. We choose the hedging instrument to be a credit swap written on the (defaultable) counterparty. Our choice is in line with current market practises. Major derivative desks routinely use credit swaps to hedge counterparty exposures (\cite{Gregory}, see Chapter 2.4), and these contracts are highly requested by market participants during periods of considerable market distress. The liquidity of credit swaps, typically higher than that of the corresponding bonds, make them better instruments to implement cost-effective hedging strategies. Hedging is only performed up to the earliest of the maturity of the portfolio and the counterparty{'}s default time, that is hedging terminates if the portfolio expires or if contingent payments are triggered by the counterparty's default.

The main conceptual novelty of our paper is the development of a comprehensive framework which simultaneously handles (i) a default intensity model enhanced with feedback from defaults (see Proposition~\ref{lem:existence}), and (ii) a dividend process for the hedging instrument (CDS) whose dynamics is of the jump-diffusive type. Earlier studies (\cite{BiaCre07}, \cite{BiaCre09}, \cite{BiaCre} and \cite{ceci2017}) consider a hedging instrument with continuous trajectories, and use an enlargement of filtration approach. The work of
{\cite{ceciEJP2015} considers hedging of a European derivative claim via default-free trading instruments following a jump-diffusion process}. Similar to \cite{FreySchimdt}, we work directly under the risk-neutral martingale (pricing) measure. As a consequence, the gain {or cumulative price} process of the CDS on the counterparty is a martingale. This setting allows us to consider risk-minimization, instead of local risk-minimization. \cite{FreySchimdt} also employ the risk-minimization approach, {but assume conditionally independent default times whose intensities depend} on an unobservable stochastic factor. Other related studies on quadratic hedging approaches to credit risk modeling include {\cite{Ok} who employ structural default models, and \cite{Wang} who consider vulnerable European contingent claims.}

There are several technical contributions in our efforts, outlined next. We propose an interacting intensity model with decaying contagion intensities, and establish its mathematical existence by constructing a sequence of solutions to piecewise stochastic differential equations (SDEs). We show that the optimal hedging strategy is given by the integrand of the GKW decomposition for the CVA payment stream (see Proposition~\ref{thm:strategy} and Theorem~\ref{thm:thetaGWK}). Such a step is fundamental for applying the risk-minimization method to our setting. We obtain an explicit formula for the risk-minimizing hedging strategy by deriving the martingale representation of the conditional expectation of the counterparty risk price payment stream (see Proposition~\ref{thm:mart-V}). This representation is in terms of unique smooth solutions to nonlinear recursive systems of Cauchy problems. These Cauchy problems are defined on an unbounded domain, have non-Lipschitz coefficients, and are linked through the default states of the economy. The nonlinearity of this system of partial differential equations (PDEs) is inherited from the nonlinear structure of the CVA. 
Our paper also makes other technical contributions related to the theory of nonlinear PDEs. Our solution approach is to prove the uniform integrability of the family generated by the Feymann-Kac's representations of the solution at any neighborhood of a fixed space-time data point. Such a property allows us to apply existence and uniqueness results from \cite{health2000} to our specific setting.


The rest of the paper is organized as follows. Section \ref{sec:model} develops the model. Section \ref{sec:cva} discusses the gain  processes and the CVA  representation. Section \ref{sec:hedge} studies the risk-minimizing CVA hedging strategy. Section \ref{sec:applications} specializes our framework to concrete portfolio credit derivatives. Section \ref{sec:conclusion} concludes. Some technical proofs are delegated to the Appendix.


\section{The Model} \label{sec:model}
We assume the existence of $N$ risky entities, referred to as name ``1'', name ``2'',..., name ``$N$''. We use ``$N+1$'' to denote the counterparty of the investor in the contract. Section \ref{sec:interact} develops the interacting default intensity model. Section \ref{sec:defclaim} develops the representation of a general defaultable claim.

\subsection{The Interacting Default Intensity Model} \label{sec:interact}
The default intensity processes are interacting jump-diffusion processes. The jumps capture the contagious impact that the default of a firm has on the default intensities of the surviving firms. Let $(\Omega,\F,\Qx)$ be a probability space endowed with a risk-neutral probability measure $\Qx$. Let $W(t) =(W_j(t))_{j=1,\ldots,K}^{\top}$, $t\geq0$, be a $K$-dimensional Brownian motion, and assume the existence of $N+1$ square-integrable positive random variables $\chi_1,\chi_2,\ldots,\chi_{N+1}$. Let $\Fx=(\F(t))_{t\geq0}$, with $\F(t)=\sigma(W(s);\ s\leq t) \vee \sigma(\chi_i ;\ i=1,\ldots,N+1)$. Denote by $H(t)=(H_1(t),\ldots,H_{N+1}(t))$ the $N+1$-dimensional default indicator process, i.e. $H_i(t) = 1$ if the {name} $i$ has defaulted before {or at} time $t$, and zero otherwise. This implies that the state space of the process $H=(H(t))_{t\geq0}$ is given by ${\cal S}:=\{0,1\}^{N+1}$. Correspondingly define the filtration $\Hx_i=({\cal H}_i(t))_{t\geq0}$ for $i=1,\ldots,N+1$, where ${\cal H}_i(t)=\sigma(H_i(s);\ s\leq t)$. The global market filtration, including default event information is given by $\Gx=(\G(t))_{t\geq0}=\Fx\vee\Hx_1\vee\cdots\vee\Hx_{N+1}$ augmented by all $\Qx$-null sets so to satisfy the usual conditions. The impact of past defaults on the default intensity of {name} $i$ is captured by the pure jump process
\begin{align}\label{eq:Li}
{J}_i(t):=\sum_{j=1}^{N +1} w_{ij}H_j(t),\qquad t\geq0.
\end{align}
The $i$-th entry of the weight vector $w_j=(w_{ij})_{i=1,\ldots,N+1}\in[0,\infty)^{N+1}$ measures the extent to which the default of {name} $i$ impacts the default intensity of {name} $j$.

Next, we introduce the interacting intensity model with decaying contagion intensities. Under the risk-neutral probability measure $\Qx$, the default intensity processes satisfy a system of interacting {SDEs} given by, for $i =1,\ldots,N+1$,
\begin{align}\label{eq:hti}
d X_i(t) &= \mu_i({X(t)})dt + \sum_{k=1}^K\sigma_{ik}\left({X(t)}\right)dW_k(t)+dJ_i(t),\quad X_{i}(0)=\chi_i.
\end{align}
and $X(t)=(X_i(t))_{i=1,\ldots,N+1}^{\top}$ for $t\geq0$. If the weight $w_{ij}$ is high, the default of {name} {$j$ increases substantially the default intensity of {name} $i$}. If $w_{ij}$'s are high for sufficiently many $i$, the probability of multiple firms defaulting within a short time after the default of {name} $j$ is high. This captures the default clustering phenomenon, empirically documented in the literature (see, for instance, \cite{GieseckeClu}). Throughout the paper, we impose the following conditions on the coefficients of Eq.~\eqref{eq:hti}:
\begin{itemize}
  \item[({\bf A1})] The coefficients $\mu(x)=(\mu_i(x))_{i=1,\ldots,N+1}^{\top}$ and $\sigma(x)=(\sigma_{ik}(x))_{i=1,\ldots,N+1;k=1,\ldots,K}$ are locally Lipchitz continuous with linear growth in $x\in\R_+^{N+1}$, $\R_+:=(0,\infty)$. Additionally, ${\rm det}((\sigma\sigma^{\top})(x))\neq0$ for $x\in\R_+^N$.
  \item[({\bf A2})] For $(t,x)\in[0,T]\times\R_+^{N+1}$, let $\tilde{X}^{t,x}(s)=(\tilde{X}_i^{t,x}(s))_{i=1,\ldots,N+1}^{\top}$ satisfy $\tilde{X}^{t,x}(t)=x$ and for $s\in[t,T]$,
      \begin{align}\label{eq:Xtilde}
      d\tilde{X}^{t,x}(s) &= \mu(\tilde{X}^{t,x}(s))ds + \sigma(\tilde{X}^{t,x}(s))d W(s).
      \end{align}
      Then it holds that $\Qx(\tilde{X}^{t,x}(s)\in\R_+^{N+1}\text{ for all }s\in[t,T])=1$.
\end{itemize}
By Theorem V.38 in \cite{Protter05}, the condition ({\bf A1}) implies that SDE \eqref{eq:Xtilde} has a unique (strong) solution, while the condition ({\bf A2}) guarantees that $\tilde{X}^{t,x}=(\tilde{X}^{t,x}(s))_{s\geq t}$ is always strictly positive if the data is strictly positive at time $t$. {Furthermore, this implies that the $i$-th default intensity process $X_i=(X_i(t))_{t\geq0}$ is strictly positive, see also Proposition \ref{lem:existence} below. The condition ${\rm det}((\sigma\sigma^{\top})(x))\neq0$ in ({\bf A1}) implies that the infinitesimal generator  of the Markov process $\tilde{X}^{t,s}$ is uniformly elliptic}, see also Lemma 3 in \cite{health2000}.


We assume that the bivariate process $(X,H)=(X(t),H(t))_{t\geq0}$ is Markovian with state space $\R_+^{N+1}\times\mathcal{S}$. Specifically, for each $i=1,\ldots,N+1$ and $t>0$, $H(t)$ transits to its neighbouring state $H^i(t):=(H_1(t),\ldots,H_{i-1}(t),1-H_i(t),H_{i+1}(t),\ldots,H_{N+1}(t))$ at the state-dependent rate ${\bf1}_{\{H_i(t)=0\}}X_i(t)$. Then the default time of the $i$-th name is given by $\tau_i:=\inf\{t>0; H_i(t)=1\}$ where $\inf\emptyset=+\infty$ by convention. Equivalently $H_i(t)={\bf1}_{\tau_i\leq t}$ for $t\geq0$.  It may be easily seen that
\begin{align}\label{eq:default-mart1C}
M_i(t) := H_i(t) - \int_{0}^{t\wedge\tau_i}X_i(s)ds,\qquad t\geq0
\end{align}
is a $\Qx$-martingale. To the best of our knowledge, the mathematical existence of this default model has not been investigated in the literature. Proposition \ref{lem:existence} establishes the existence of such a process. 
\begin{prop}\label{lem:existence}
Under assumptions ({\bf A1}) and ({\bf A2}), there exists a unique $\R_+^{N+1}\times{\cal S}$-valued and $\Gx$-adapted Markov process $(X,H)$ satisfying \eqref{eq:Li}, \eqref{eq:hti} and \eqref{eq:default-mart1C}.
\end{prop}

\noindent{\it Proof.}\quad We first introduce convenient notation. We use $z=0^{j_1,\ldots,j_l}$ to denote the vector obtained by flipping the entries $j_1 \neq j_2,\cdots\neq j_l$ of the zero vector to one. Clearly, $0^{j_1,\ldots,j_{N+1}}={e_{N+1}}$  (here $e_{N+1}$ denotes the canonical row vector with all entries equal to one). We construct $(X(t),H(t))$ for $t\geq0$ iteratively. More precisely, for $i=1,\ldots,N+1$, we first consider the following SDE given by
\begin{align}\label{eq:SDE5}
\D X_i^{(0)}(t)= \mu_i(X^{(0)}(t))\D t + \sum_{k=1}^K\sigma_{ik}(X^{(0)}(t))\D W_k(t),\quad X_i^{(0)}(0)= \chi_i>0.
\end{align}
{Here $X^{(0)}(t)=(X_i^{(0)}(t))_{i=1,\ldots,N+1}^{\top}$ for $t\geq0$.}
The assumptions ({\bf A1}) and ({\bf A2}) imply that SDE~\eqref{eq:SDE5} admits a unique strong solution which is strictly positive for each $i=1,\ldots,N+1$.

Let $({\xi}_{ij};\ i,j=1,\ldots,N+1)$ be independent standard exponentially distributed random variables, which are also independent of the $K$-dimensional Brownian motion. At the initial time, no default occurs, i.e., ${H}(0)={0}$. Then the first time when any of $N+1$ names defaults is denoted by $\hat\tau_1$, and given by
\begin{align*}
\hat\tau_1&:=\min_{i=1,\ldots,N+1}\tau_{1i},\nonumber\\
\tau_{1i}&:=\inf\left\{t>0;\ \int_0^tX_i^{(0)}(u)\D u\geq\xi_{1i}\right\},\ \ \ \ i=1,\ldots,N+1.
\end{align*}
For $i=1,\ldots,N+1$, we set $X_i(u)=X_i^{(0)}(u)$ and $H(u)=H(0)=0$ when $u\in[0,\hat\tau_1)$. Further, define $i_1:=\mathop{\arg\min}\limits_{i=1,\ldots,N+1}\tau_{1i}$ and let $X_{i_1}^{(1)}(t)=0$ for $t\geq\hat{\tau}_1$. For $i\in\{1,\ldots,N+1\}\setminus\{i_1\}$, consider the following SDE: on $t\geq\hat{\tau}_1$,
\begin{align}\label{eq:SDE6}
X_i^{(1)}(t)&= X_i^{(0)}(\hat{\tau}_1)+\int_{\hat{\tau}_1}^t\mu_i(X^{(1)}(u))\D u + \sum_{k=1}^K\int_{\hat{\tau}_1}^t\sigma_{ik}(X^{(1)}(u))\D W_k^{(1)}(u)+w_{ii_1}.
\end{align}
Here $W_k^{(1)}(t):=W_k(t+\hat{\tau}_1)-W_k(\hat{\tau}_1)$ and { $X^{(1)}(t)=(X_i^{(1)}(t))_{i=1,\ldots,N+1}^{\top}$} for $t\geq0$. The assumptions ({\bf A1}) and ({\bf A2}) imply that Eq.~\eqref{eq:SDE6} admits a unique positive strong solution $X_i^{(1)}(t)$ on $t\geq\hat\tau_1$ since $w_{ii_1}>0$. Furthermore we define the second default time as
\begin{align*}
\hat\tau_2&:=\min_{i\in\{1,\ldots,N+1\}\setminus \{i_1\}}\tau_{2i},\nonumber\\
\tau_{2i}&:=\inf\left\{t\geq\hat\tau_1;\ \int_{\hat\tau_1}^tX_i^{(1)}(u)\D u\geq\xi_{2i}\right\},\ \ \ \ i\in\{1,\ldots,N+1\}\setminus \{i_1\}.
\end{align*}

Similarly to the above construction of $(X(t),H(t))$ on $t\in[0,\hat\tau_1)$,  for $u\in[\hat\tau_1,\hat\tau_2)$, we set $X_i(u)=X_i^{(1)}(u)$ for all $i\in\{1,\ldots,N+1\}\setminus\{i_1\}$, and ${H}(u)={H}(\hat\tau_1)={0^{i_1}}$.
Moreover, set $i_2:=\mathop{\arg\min}\limits_{i\in\{1,\ldots,N+1\}\setminus\{i_1\}}\tau_{2i}$. More generally, for $n=3,\ldots,N$, the $n$-th default time is specified by
\begin{align*}
\hat\tau_n&:=\min_{i\in\{1,\ldots,N+1\}\setminus \{i_1,\ldots,i_{n-1}\}}\tau_{ni},\nonumber\\
 \tau_{ni}&:=\inf\left\{t\geq\hat\tau_{n-1};\ \int_{\hat\tau_{n-1}}^tX_i^{(n-1)}(u)\D u\geq\xi_{ni}\right\},\ \ i\in\{1,\ldots,N+1\}\setminus \{i_1,\ldots,i_{n-1}\}.
\end{align*}
Above $i_1,\ldots,i_{n-1}$ are defined in a similar way to $i_1$ and $i_2$ following a recursive process. For $i\in\{1,\ldots,N+1\}\setminus\{i_1,\ldots,i_{n-1}\}$, and for $t\geq\hat{\tau}_{n-1}$,
\begin{align}\label{eq:SDE8}
X_i^{(n-1)}(t)&=X_i^{(n-2)}(\hat{\tau}_{n-1})+\int_{\hat{\tau}_{n-1}}^t\mu_i(X^{(n-1)}(u))\D u\nonumber\\
&\quad + \sum_{k=1}^K\int_{\hat{\tau}_{n-1}}^t\sigma_{ik}(X^{(n-1)}(u))\D W_k^{(n-1)}(u)+\sum_{j\in\{i_1,\ldots,i_{n-1}\}}w_{ij}
\end{align}
with $W_k^{(n-1)}(t):=W_k(t+\hat{\tau}_{n-1})-W_k(\hat{\tau}_{n-1})$ and  {$X^{(n-1)}(t)=(X_i^{(n-1)}(t))_{i=1,\ldots,N+1}^{\top}$} for $t\geq0$. The assumptions ({\bf A1}) and ({\bf A2}) imply that Eq.~\eqref{eq:SDE8} admits a unique positive solution since $\sum_{j\in\{i_1,\ldots,i_{n-1}\}}w_{ij}>0$. We can repeat the above recursive procedures and establish the Markov process $(X(t),H(t))$ on $t\in[\hat\tau_{n-1},\hat\tau_n)$ until $n=N+1$. If $t\geq\hat\tau_{N+1}$, all names in the pool have defaulted. Using the argument given in Section~4 of \cite{Lando98}, we conclude that $(X(t),H(t))_{t\geq0}$ is the desired Markov process. \hfill$\Box$\\

\subsection{Defaultable Claims} \label{sec:defclaim}

We introduce the formalism to describe the class of defaultable claims treated in this paper. The specification is general enough to accommodate a large class of portfolio credit derivatives, of which the credit valuation adjustment can be computed.
\begin{definition}\label{def:quadruple}
{Let $\xi(z)$, $a(z)$, $Z(z)$ and $K(z)$, $z \in {\cal S}$, be measurable functions}. A defaultable claim maturing at $T>0$ is a quadruple $(\xi,a,Z,K)$, where the random variable $\xi:=\xi(H(T))$, the processes $a(t):=a(H(t))$ and $Z(t):=Z(H(t))$ for $t\in[0,T]$. The process $K(t):=K(H(t))$, $t\in[0,T]$, is the indicator function of a positive $\Gx$-stopping time $\bar{\tau}$, i.e. it holds that $K(t)={\bf1}_{\bar{\tau}\leq t}$. 
\end{definition}
The financial meaning of the components of a defaultable claim becomes clear from the definition of the dividend, or total cash flow, process. Such a process describes all cash flows generated by the defaultable claim over its lifespan $(0,T]$, that is, after the contract was initiated at time 0. Hereafter, we introduce the following notations
\begin{align}\label{eq:trasferdefault}
a^j(t):=a(H^j(t)),\quad Z^j(t):=Z(H^j(t)),\quad K^{j}(t):= K(H^j(t)),
\end{align}
where for $z\in{\cal S}$,
\begin{align}\label{eq:flip-vec}
z^j:=(z_1,\ldots,z_{j-1},1-z_j,z_{j+1},\ldots,z_{N+1}),\qquad j=1,\ldots,N+1
\end{align}
is obtained by flipping the $j$-th component of $z$ from zero to one, or vice versa.

\begin{definition}\label{def:dividend}
The dividend processes $D = (D(t))_{t\geq 0}$ associated with the defaultable claim $(\xi,a,Z,K)$ maturing at $T$ equals, for every $t \geq 0$,
\[
D(t) = \xi (1-K(T)) {\bf1}_{[T, \infty)}(t) + \int_{0}^{t \wedge T} (1-K(u)) a(u) du + \int_{0}^{t \wedge T} Z(u) d K(u).
\]
\end{definition}
It is clear from the above definition that the process $D=(D(t))_{t\geq0}$ has finite variation. It admits the following financial interpretation: {the r.v.} $\xi$ is the promised payoff, $a=(a(t))_{t\geq0}$ represents the process of promised dividends and the process $ Z:=(Z(t))_{t\geq0}$ specifies the payoff delivered when the indicator process $K=(K(t))_{t\geq0}$ changes from zero to one.  {Notice that we allow for quantities to depend on the default process $H$, and that the process $Z$ is not assumed to be $\Gx$-predictable.} 
Such a setup differs from earlier works, see for instance \cite{BielAAP}, and allows us to use the same general
framework to hedge counterparty risk of a larger set of defaulable claims, including those whose recovery process
depends on a totally inaccessible stopping time.

\subsection{Examples}\label{sec:example}
The proposed framework can be specialized to deal with a class of credit derivatives, which are routinely used by investors to hedge risks. Without loss of generality, assume that the notional amount of the considered contracts is one.

\noindent{\bf Default Intensities.}\quad {For $i=1,\ldots,N+1$, assume that the default intensity of the $i$-th {reference entity} follows the dynamics}
\begin{align}\label{eq:htiCIR}
\D X_i(t) &= (\kappa_i-\nu_iX_i(t))\D t + \sum_{k=1}^K\sigma_{k}\sqrt{X_i(t)}\D W_k(t)+d{J}_i(t),\quad X_i(0)=\chi_i>0.
\end{align}
The parameters $\kappa_i,\nu_i$, $i=1,\ldots,N+1$, and $\sigma_k$, $k=1,\ldots,K$, are positive constants satisfying the following Feller's boundary classification condition: $2\kappa_i\geq\sum_{k=1}^K\sigma_{k}^2$, for $i=1,\ldots,N+1$.
This implies that the assumption ({\bf A2}) holds. The default intensity mean reverts to its long-run level given by $\frac{\kappa_j}{\nu_j}>0$ between two consecutive default events. This captures the empirically observed time decaying effect of default intensities. When a firm $i$ defaults, the default intensity of firm $j$ instantaneously jumps upward. The contagion effect decays at an exponential rate.\\


\noindent{\bf Credit Swap Portfolio.}\quad
Consider a portfolio of credit default swap contracts whose reference entities are denoted by ``$1$'', ``$2$'', ..., ``$N$'', and recall that the counterparty of the investor is denoted by ``$N+1$''. In a credit default swap contract, the protection leg commits to paying a contractually specified spread premium $\varepsilon_i>0$ until the earliest of the default time $\tau_i$ of the reference entity or the maturity $T$ of the contract. The protection seller pays the loss rate $L_i(t):=L_i(H(t))\in(0,1]$ times the given notional amount at the time $\tau_i$ that the $i$-th reference entity defaults. This loss rate may depend on the default state of the portfolio.

Consider the case that all credit default swaps have the same maturity $T>0$, and we view the payoff from the point of view of the protection seller. The quadruple $(\xi_i,a_i,Z_i,K_i)$ for $i=1,\ldots,N+1$, is specified as follows:
$$
\xi_i = 0, \qquad\; a_i(t) = -\varepsilon_i, \qquad\; Z_i(t) = L_i(t), \qquad\; K_i(t) = H_i(t),
$$
i.e. $K_i(t) = H_i(t)$ is the indicator of the default time of the $i$-th reference entity ($\bar{\tau}_i=\tau_i$). From Definition~\ref{def:dividend}, the representation of the dividend process of the $i$-th CDS is given by
\begin{align}\label{eq:dividend-cds}
D_i(t) &= -\varepsilon_i\int_{0}^{t \wedge T} (1-H_i(u))du + \int_{0}^{t \wedge T} L_i(u) d H_i(u)\nonumber\\
&=-\varepsilon_i(t \wedge T\wedge\tau_i) + L_i(\tau_i){\bf1}_{\tau_i\leq t \wedge T}.
\end{align}

\noindent{\bf Risky Bond Portfolio.}\quad
Consider a portfolio of coupon paying bonds underwritten by firms ``$1$'', ``$2$'', ..., ``$N$''. The seller of the bond of firm $i$ receives the promised coupon payments $\varepsilon_i > 0$ until the earliest of maturity or default of firm $i$. If the firm $i$ has not defaulted by $T$, then the seller also receives a notional payment. If the firm $i$ defaults before the maturity $T$, the owner of the bond receives the recovery rate $R_i(t) := 1-L_i(H(t))\in[0,1)$ at the time $\tau_i$ that firm $i$ defaults. This recovery rate may depend on the default state of the portfolio. Then we have that the quadruple $(\xi_i,a_i,Z_i,K_i)$ for $i=1,\ldots,N$, is specified as follows:
$$
\xi_i = 1, \qquad \; a_i(t) = \varepsilon_i, \qquad \; Z_i(t) = R_i(t)=1-L_i(t), \qquad \; K_i(t) = H_i(t),
$$
i.e. $K_i(t) = H_i(t)$ is the indicator of the default time of the $i$-th reference entity ($\bar{\tau}_i=\tau_i$).
Following Definition~\ref{def:dividend}, the representation of the dividend process of the $i$-th risky bond is given by
\begin{align}\label{eq:dividend-bond}
D_i(t) &= (1-H_i(T)){\bf1}_{t\geq T}+\varepsilon_i\int_{0}^{t \wedge T} (1-H_i(u))du + \int_{0}^{t \wedge T} R_i(u) d H_i(u)\nonumber\\
&=(1-H_i(T)){\bf1}_{t\geq T}+\varepsilon_i(t \wedge T\wedge\tau_i) + R_i(\tau_i){\bf1}_{\tau_i\leq t\wedge T}.
\end{align}

\noindent{\bf First-to-Default Claim.}\quad
In a first-to-default swap, the protection buyer will make the spread premium payment $\varepsilon > 0$ to the protection seller. The protection seller, in return, will be required to pay the loss rates times the given notational to the protection buyer if and when any one of the reference entities ``$1$'', $\ldots$, ``$N$'' defaults before the contract expires at $T$. The payment will only be made for the first entity to default, i.e. the payment will be $L_i(t):=L_i(H(t))\in(0,1]$ if $i$ is the first entity to default. This deal is typically executed by a firm which wants to hedge its exposure to a number of different firms.

Assume that the notional amount is one, and let view the payoff from the point of view of the protection seller. Then we have that the quadruple $(\xi,a,Z,K)$ is specified as follows:
$$
\xi= 0, \qquad\; a(t) = -\varepsilon, \qquad\; Z(t) =\sum_{i=1}^N L_i(t)H_i(t), \qquad\; K(t) = 1- \prod_{i=1}^N (1-H_i(t)),
$$
i.e. $\bar{\tau}_1=\tau_1\wedge\cdots\tau_N$.

\begin{lemma}\label{lem:Dfirst-todefault}
The dividend process of the first-to-default claim admits the representation given by
\begin{align}\label{eq:Dfirst}
D(t) 
&=-\varepsilon (t\wedge T\wedge\bar{\tau}_1)+\sum_{i=1}^NL_i(\bar{\tau}_1){\bf1}_{\tau_i=\bar{\tau}_1}{\bf1}_{\bar{\tau}_1\leq t\wedge T},
\end{align}
where $\bar{\tau}_1=\tau_1\wedge\cdots\tau_N$ is the first-to-default time.
\end{lemma}

\section{Gain Processes and CVA Representation}\label{sec:cva}
In this section, we study hedging of counterparty risk for a general defaultable claim, including portfolio credit derivatives. The hedging instrument is a credit default swap referencing the risky counterparty ``$N$+1''. Throughout the paper, we set the interest rate to zero. Such an assumption allows us to avoid unnecessary clutter of notation, and to highlight the main probabilistic forces. The whole analysis can be generalized in a straightforward fashion to the case of nonzero interest rate.

The establishment of the hedging framework consists of the following steps. Section \ref{sec:pricegain} studies the price representation and the dynamics of the gain process {(or cumulative price process)} for a defaultable claim specified in Definition~\ref{def:quadruple}.
Section \ref{sec:CVApricerep} characterizes the credit valuation adjustment (CVA) of the claim, and derives the representation of the stopped payment stream associated with the CVA.

\subsection{Price and Gain Processes} \label{sec:pricegain}

Let $(\xi,a,Z,K)$ be a defaultable claim as in Definition~\ref{def:quadruple}. 
For any fixed time $t \in [0,T]$, the process $(D(u)-D(t))_{u\in[t,T]}$ represents all cash flows generated by the defaultable claim $(\xi,a,Z,K)$ in the interval $[t,T]$. Such a process may depend on the past behavior of the claim as well as on the history of the market prior to time $t$. Clearly, the past cash flows are not valued by the market, so that the market value at time $t$ of a defaultable claim only reflects future cash flows to be paid/received over the time interval $(t,T]$.

The price process $(S(t,T))_{t\in[0,T]}$ of the defaultable claim $(\xi,a,Z,K)$ equals $Z(\bar{\tau})$ at the default time $\bar{\tau}$, and zero after the default, that is $S(t,T)=0$ on $\{ t > \bar{\tau}\}$. On $\{ \bar{\tau} > t \}$,  the pre-default price is given by its risk-neutral expected payoff of dividend payments, i.e., for $t\in[0,T]$,
\begin{align}\label{eq:StT}
{S}(t,T) = \Ex\left[D(T) - D(t) \big|\G_t\right].
\end{align}
Above, $\Ex$ denotes the expectation under the pricing measure $\Qx$. Correspondingly the gain process of the defaultable claim $(\xi,a,Z,K)$ (see also \cite{FreySchimdt} for a related definition) as, for $t\in[0,T]$,
\begin{align}\label{eq:gain}
Y(t):=\Ex[D(T)|\G_t].
\end{align}
Notice that $Y(t)=S(t,T)+D(t)$ on $\{\bar{\tau}>t\}$, i.e. the gain process  is given by the sum of the current market value and the dividend payments. By virtue of Definition~\ref{def:dividend}, the dividend process of the defaultable claim $(\xi,a,Z,K)$ is given by, for $t\in[0,T]$,
\begin{align}\label{eq:dividendT}
D(t) = \xi(1-K(T)){\bf1}_{t=T} +\int_{0}^{t} (1-K(u)) a(u) du + \int_{0}^{t} Z(u) d K(u).
\end{align}

We next study the representation of the time-$t$ price $S(t,T)$ given by \eqref{eq:StT}, which will be used to characterize the CVA representation of the portfolio of defaultable claims in the following subsection.
\begin{prop}\label{lem:pre-default}
For $t\in[0,T]$, the {time $t$} price $S(t,T)$ given by \eqref{eq:StT} admits the following representation:
\begin{align}\label{eq:StT2}
S(t,T)={\bf1}_{t\neq T}{\Lambda}_1(t,X(t),H(t))+{\Lambda}_2(t,X(t),H(t))-Z(t)K(t),
\end{align}
where, for $(t,x,z)\in[0,T]\times\R_+^{N+1}\times{\cal S}$,
\begin{align}\label{eq:F1F2}
{\Lambda}_1(t,x,z)&:=\Ex_{t,x,z}\left[\xi(1-K(T))\right],\nonumber\\
{\Lambda}_2(t,x,z)&:=\Ex_{t,x,z}\bigg[Z(T)K(T)+\int_{t}^{T} (1-K(u)) a(u) du\\
&\quad-\sum_{j=1}^{N+1}\int_t^T K(u)[Z^j(u)-Z(u)](1-H_j(u))X_j(u)du\bigg].\nonumber
\end{align}
Above, we have used the abbreviation $\Ex_{t,x,z}[\cdot]:=\Ex[\cdot|X(t)=x,H(t)=z]$ for the conditional expectation, and we recall that $Z^j(u)$ has been defined in Eq.~\eqref{eq:flip-vec}.
\end{prop}
\noindent{\it Proof.}\quad Using Eq.~\eqref{eq:dividendT}, it holds that, for $t\in[0,T]$,
\begin{align*}
D(T)-D(t)&=\xi(H(T))(1-K(T)){\bf1}_{t\neq T}+\int_{t}^{T} (1-K(u)) a(u) du + \int_{t}^{T} Z(u) d K(u).
\end{align*}
Then it follows from Eq.~\eqref{eq:StT} that, for $t\in[0,T]$,
\begin{align*}
{S}(t,T)&=\Ex\Bigg[ \xi(H(T))(1-K(T)){\bf1}_{t\neq T}+ \int_{t}^{T} (1-K(H(u))) a(H(u)) du\nonumber\\
&\quad + \int_{t}^{T} Z(H(u)) d K(H(u))\Big|\G_t\Bigg].
\end{align*}
Notice that $(Z(H(t)))_{t\in[0,T]}$ and $(K(H(t)))_{t\in[0,T]}$ are pure jump processes. Using integrations by parts, it follows that
\begin{align}\label{eq:inteparts}
Z(H(T))K(H(T))=&Z(H(t))K(H(t))+\int_t^T Z(H(u))dK(H(u))\nonumber\\
&+\int_t^T K(H(u^-))dZ(H(u)).
\end{align}
On the other hand, the It\^o's formula gives that for $u\in[t,T]$,
\begin{align*}
dZ(H(u))&=\sum_{j=1}^{N+1} [Z(H^j(u^-))-Z(H(u^-))]dH_j(u)=\sum_{j=1}^{N+1} [Z(H^j(u^-))-Z(H(u^-))]dM_j(u)\nonumber\\
&\quad+\sum_{j=1}^{N+1}[Z(H^j(u))-Z(H(u))](1-H_j(u))X_j(u)du.
\end{align*}
For $j=1,\ldots,N+1$, recall that $M_j=(M_j(t))_{t\in[0,T]}$ is the $\Gx$-martingale given by Eq.~\eqref{eq:default-mart1C}.
Hence, the equality~\eqref{eq:inteparts} yields that
\begin{align*}
&\int_t^T Z(H(u))dK(H(u))=Z(H(T))K(H(T))-Z(H(t))K(H(t))-\int_t^T K(H(u^-))dZ(H(u))\nonumber\\
&\qquad\qquad=Z(H(T))K(H(T))-Z(H(t))K(H(t))\nonumber\\
&\qquad\qquad\quad-\sum_{j=1}^{N+1}\int_t^T K(H(u^-))[Z(H^j(u^-))-Z(H(u^-))]dM_j(u)\nonumber\\
&\qquad\qquad\quad-\sum_{j=1}^{N+1}\int_t^TK(H(u))[Z(H^j(u))-Z(H(u))](1-H_j(u))X_j(u)du.
\end{align*}
This results in the price representation given by $S(t,T)=F(t,X(t),H(t))-Z(H(t))K(H(t))$, where
\begin{align}\label{eq:F}
F(t,x,z)&:=\Ex_{t,x,z}\bigg[ \xi(H(T))(1-K(H(T))){\bf1}_{t\neq T}+ Z(H(T))K(H(T))+\int_{t}^{T} (1-K(H(u))) a(H(u)) du\nonumber\\
&\quad-\sum_{j=1}^{N+1}\int_t^T K(H(u))[Z(H^j(u))-Z(H(u))](1-H_j(u))X_j(u)du\bigg],
\end{align}
using that the pair $(X,H)$ is a $\Gx$-adapted Markov process. Then the price representation \eqref{eq:StT2} follows from the decomposition of the price function $F(t,x,z)$ given by
\begin{align}\label{eq:F-decom}
F(t,x,z)={\bf1}_{t\neq T} \Lambda_1(t,x,z) + \Lambda_2(t,x,z),\qquad (t,x,z)\in[0,T]\times\R_+^{N+1}\times{\cal S}.
\end{align}
This completes the proof of the lemma.\hfill$\Box$\\

Next, we characterize the functions $\Lambda_1$ and $\Lambda_2$ given in \eqref{eq:F1F2}, and further study the dynamics of the gain process $Y=(Y(t))_{t\in[0,T]}$ of the defaultable claim $(\xi,a,Z,K)$ given by Eq.~\eqref{eq:gain}. To this purpose, for $\alpha=(\alpha_1,\alpha_2,\alpha_3)\in\R^3$,
consider the following recursive system of backward Cauchy problems given by, on $(t,x,z)\in[0,T)\times\R_+^{N+1}\times{\cal S}$,
\begin{align}\label{eq:Cahchypro1}
&\left(\frac{\partial}{\partial t} + \A\right)F_{\alpha}(t,x,z)+\alpha_3(1-K(z))a(z)-\alpha_3\sum_{j=1}^{N+1}K(z)[Z(z^j)-Z(z)](1-z_j)x_j= 0
\end{align}
with terminal condition
\begin{align}\label{eq:terminal-condition}
F_{\alpha}(T,x,z)=\alpha_1\xi(z)(1-K(z))+\alpha_2Z(z)K(z),\qquad (x,z)\in \R_+^{N+1}\times{\cal S}.
\end{align}
In the above expression, the operator ${\cal A}$ is the generator of the Markov process $(X, H)$. It is a difference-differential operator acting on the smooth function $f(\cdot,z)$ for each $z\in{\cal S}$, and is given by
\begin{align}\label{eq:A}
{\cal A}f(x,z) &:=\tilde{\cal A}f(x,z)+\sum_{j=1}^{N+1} \big[f(x+w_j,z^j) - f(x,z)\big](1-z_j)x_j,
\end{align}
where the vector of weights $w_j=(w_{ij})_{i=1,\ldots,{N+1}}$, {and recall that the default state $z^j$ has been defined in Eq.~\eqref{eq:flip-vec}}. The second-order differential operator $\tilde{\cal A}$ is defined by, for $(x,z)\in\R_+^{N+1}\times{\cal S}$,
\begin{align}\label{eq:tildeA}
\tilde{\cal A}f(x,z) :=\mu(x)^{\top}D_xf(x,z)+\frac{1}{2}{\rm tr}[(\sigma\sigma^{\top})(x)D_x^2f(x,z)],
\end{align}
and is uniformly elliptic under the assumption ({\bf A2}). In terms of Eq.~\eqref{eq:A}, we can rewrite the Cauchy problem~\eqref{eq:Cahchypro1} in the following equivalent form:
\begin{align}\label{eq:Cahchypro2}
0&=\left(\frac{\partial}{\partial t} + \tilde{\A}\right)F_{\alpha}(t,x,z)+\alpha_3(1-K(z))a(z)-\alpha_3\sum_{j=1}^{N+1}K(z)[Z(z^j)-Z(z)](1-z_j)x_j\nonumber\\
&\qquad+\sum_{j=1}^{N+1}\big[F_{\alpha}(t,x+w_j,z^j)-F_{\alpha}(t,x,z)\big](1-z_j)x_j.
\end{align}

We next illustrate the recursive structure of the system of backward Cauchy problems \eqref{eq:Cahchypro2} in terms of default states $z\in{\cal S}$. Recall that $z=0^{j_1,\ldots,j_l}$ denotes the vector with zero entries except for the components $j_1 \neq j_2,\cdots\neq j_l$ which are set to one. Clearly, $0^{j_1,\ldots,j_{N+1}}={e_{N+1}}$  (here $e_{N+1}$ denotes the canonical row vector with all entries equal to one). For any measurable function $f(t,x,z)$, defined on $(t,x,z)\in[0,T]\times\R_+^{N+1}\times{\cal S}$,  for $l=0,1,\ldots,N+1$, set
\begin{align}\label{eq:flfl1i}
f^{(l)}(t,x):=f(t,x,0^{j_1,\ldots,j_{l}}),\qquad f^{(l+1),i}(t,x):=f(t,x,0^{j_1,\ldots,j_l,i}),\ i\notin\{j_1,\ldots,j_l\}.
\end{align}
We also set $f^{(0)}(t,x):=f(t,x,0)$. We distinguish two cases:
\begin{itemize}
\item $l=N+1$, i.e. all names have defaulted. In this case, the Cauchy problem~\eqref{eq:Cahchypro2} is reduced to
\begin{align}\label{eq:CahchyprolN+1}
\left(\frac{\partial}{\partial t} + \tilde{\A}\right)F_{\alpha}^{(N+1)}(t,x)+\alpha_3(1-K^{(N+1)})a^{(N+1)}=0
\end{align}
with terminal condition $F_{\alpha}^{(N+1)}(T,x)=\alpha_1\xi^{(N+1)}(1-K^{(N+1)})+\alpha_2Z^{(N+1)}K^{(N+1)}$ for all $x\in \R_+^{N+1}$. It can be easily seen that the solution admits the closed-form representation given by, for $(t,x)\in[0,T]\times\R_+^{N+1}$,
\begin{align}\label{eq:FeN+1}
F_{\alpha}^{(N+1)}(t,x)&=\alpha_1{\xi^{(N+1)}(1-K^{(N+1)})}+\alpha_2Z^{(N+1)}K^{(N+1)}+\alpha_3(1-K^{(N+1)})a^{(N+1)}(T-t).
\end{align}


\item $0\leq l\leq N$, i.e. the names $j_1,\ldots,j_l$ have defaulted. Then the Cauchy problem~\eqref{eq:Cahchypro2} becomes
\begin{align}\label{eq:Cahchyprol}
0&=\left(\frac{\partial}{\partial t} + \tilde{\A}\right)F_{\alpha}^{(l)}(t,x)-\left(\sum_{j\notin\{j_1,\ldots,j_l\}}x_j\right) F_{\alpha}^{(l)}(t,x)+\alpha_3(1-K^{(l)})a^{(l)}\nonumber\\
&\quad-\alpha_3\sum_{j\notin\{j_1,\ldots,j_l\}}K^{(l)}[Z^{(l+1),j}-Z^{(l)}]x_j+\sum_{j\notin\{j_1,\ldots,j_l\}}F_{\alpha}^{(l+1),j}(t,x+w_j)x_j.
\end{align}
The terminal condition is given by $F_{\alpha}^{(l)}(T,x)=\alpha_1\xi^{(l)}(1-K^{(l)})+\alpha_2Z^{(l)}K^{(l)}$ for all $x\in \R_+^{N+1}$. Notice that $F_{\alpha}^{(l+1),j}(t,x)=F_{\alpha}^{(N+1)}(t,x)$ given in \eqref{eq:FeN+1} if $l=N$.
\end{itemize}

We next prove that the Cauchy problem \eqref{eq:Cahchyprol} has a unique bounded classical solution $F_{\alpha}^{(l)}(t,x)$ if the Cauchy problem \eqref{eq:Cahchypro2} admits a unique bounded classical solution $F_{\alpha}^{(l+1),j}(t,x)$ if $z=0^{j_1,\ldots,j_l,j}$ for $j\notin\{j_1,\ldots,j_l\}$. The main result is stated in the following proposition {whose proof is postponed to the Appendix.}
\begin{prop}\label{lem:sol-Fi2}
Let assumptions ({\bf A1}) and ({\bf A2}) hold. Assume that at the default state $z=0^{j_1,\ldots,j_l,j}$ for $j\notin\{j_1,\ldots,j_l\}$, the Cauchy problem~\eqref{eq:Cahchypro2} admits a unique bounded classical solution $F_{\alpha}^{(l+1),j}(t,x)$ on $[0,T]\times\R_+^{N+1}$. Then at the default state $z=0^{j_1,\ldots,j_l}$, the Cauchy problem~\eqref{eq:Cahchypro2} admits a unique bounded classical solution $F_{\alpha}^{(l)}(t,x)$ on $[0,T]\times\R_+^{N+1}$. Moreover the solution admits the following recursive representation given by
\begin{align}\label{eq:FY-Fi2}
F_{\alpha}^{(l)}(t,x)&=\big(\alpha_1\xi^{(l)}(1-K^{(l)})+\alpha_2Z^{(l)}K^{(l)}\big)\Ex\left[e^{-\int_t^T\big(\sum_{k\notin\{j_1,\ldots,j_l\}}\tilde{X}_k^{(t,x)}(u)\big)du}\right] \nonumber\\ &\quad+\alpha_3(1-K^{(l)})a^{(l)}\Ex\left[\int_t^T e^{-\int_t^s\big(\sum_{k\notin\{j_1,\ldots,j_l\}}\tilde{X}_k^{(t,x)}(u)\big)du}ds\right]\\
&\quad+\sum_{j\notin\{j_1,\ldots,j_l\}}\Ex\Bigg[\int_t^T\tilde{X}_j^{(t,x)}(s)
\big[F_{\alpha}^{(l+1),j}(s,\tilde{X}^{(t,x)}(s)+w_j)-\alpha_3K^{(l)}(Z^{(l+1),j}-Z^{(l)})\big]\nonumber\\
&\qquad\qquad\times e^{-\int_t^s\big(\sum_{k\notin\{j_1,\ldots,j_l\}}\tilde{X}_k^{(t,x)}(u)\big)du}ds\Bigg].\nonumber
\end{align}
Above, the underlying $\R_+^{N+1}$-valued process $(\tilde{X}^{(t,x)}(s))_{s\in[t,T]}$ is the unique strong solution of SDE~\eqref{eq:Xtilde}.
\end{prop}

Using Proposition~\ref{lem:sol-Fi2} above and the Feynman-Kac's formula, the functions $\Lambda_1(t,x,z)$ and $\Lambda_2(t,x,z)$ defined in \eqref{eq:F1F2} can be identified as:
\begin{align}\label{eq:FalphaF12}
\Lambda_1(t,x,z) = F_{(1,0,0)}(t,x,z),\qquad \Lambda_2(t,x,z) = F_{(0,1,1)}(t,x,z).
\end{align}
The dynamics of the gain process $Y=(Y(t))_{t\in[0,T]}$ of the defaultable claim $(\xi,a,Z,K)$ can be easily obtained from Proposition~\ref{lem:sol-Fi2}. {The proof is reported in the Appendix.}
\begin{lemma}\label{lem:gain}
Suppose that assumptions ({\bf A1}) and ({\bf A2}) hold. The gain process defined by Eq.~\eqref{eq:gain} satisfy the following dynamics, for $t\in[t,T]$,
\begin{align}\label{eq:Yi}
dY(t) &= V(t,X(t),H(t))^{\top} \sigma(X(t))\D W(t)\\
&\quad+  \sum_{j=1}^{N+1}\big\{G_{j}(t,X(t^-),H(t^-))-K(t^-){[Z^j(t^-)-Z(t^-)]}\big\}\D M_j(t).\nonumber
\end{align}
For $j=1,\ldots,N+1$, $M_j=(M_j(t))_{t\in[0,T]}$ is the $\Gx$-default martingale given by Eq.~\eqref{eq:default-mart1C}. For $(t,x,z)\in[0,T]\times\R_+^{N+1}\times{\cal S}$,
\begin{align}\label{eq:VG}
V(t,x,z)&:=D_xF_{(1,1,1)}(t,x,z),\qquad
G_{j}(t,x,z):=F_{(1,1,1)}(t,x+w_j,z^j) - F_{(1,1,1)}(t,x,z),
\end{align}
for $j=1,\ldots,N+1$. The function $F_{(1,1,1)}(t,x,z)$ is the unique classical solution to the recursive system of Cauchy problems \eqref{eq:Cahchypro1} and \eqref{eq:terminal-condition}, in which we set $\alpha=(1,1,1)$. We use $D_xF_{(1,1,1)}(t,x,z)$ to denote the gradient (column) vector of $F_{(1,1,1)}(t,x,z)$ w.r.t. $x\in\R_+^{N+1}$.
\end{lemma}

\subsection{Credit Valuation Adjustment (CVA)}\label{sec:CVApricerep}

The credit valuation adjustment (CVA) is the market price of counterparty risk, see \cite{Brigoetal14,capmig}. Our goal is to compute the dynamic hedge of the credit valuation adjustment of a portfolio consisting of a finite number of defaultable claims of the form given in Definition~\ref{def:quadruple}.
\begin{definition}\label{def:port-def}
Let $\bar{N}\geq1$. For each $i=1,\ldots,\bar{N}+1$, let $(\xi_i,a_i,Z_i,K_i)$ be a defaultable claim as in Definition~\ref{def:quadruple}, where $K_i(t)=K_i(H(t))={\bf1}_{\bar{\tau}_i\leq t}$ for $t\in[0,T]$, and $\bar{\tau}_i$'s, $i=1,\ldots,N+1$, are positive $\Gx$-stopping times such that $K_1(t),\ldots,K_{\bar{N}+1}(t)$ do not jump simultaneously. We call $(\xi_i,a_i,Z_i,K_i)_{i=1,\ldots,\bar{N}+1}$ a defaultable claim portfolio.
\end{definition}

\begin{example}\label{exam:portfolio}
We provide concrete examples of a defaultable claim portfolio $(\xi_i,a_i,Z_i,K_i)_{i=1,\ldots,\bar{N}+1}$ composed of the claims considered in Example~\ref{sec:example}.

{\bf CDS portfolio.}\quad  For $i=1,\ldots,N+1$, we have
 \begin{align}\label{eq:CDS}
  \xi_i=0,\quad a_i(t)=-\varepsilon_i,\quad Z_i(t)=L_i(t),\quad K_i(t)=H_i(t).
 \end{align}
In terms of the Definition~\ref{def:port-def}, we have that $\bar{N}=N$, and $\bar{\tau}_i=\tau_i$ for $i=1,\ldots,N+1$.

{\bf Risky bonds portfolio.}\quad For $i=1,\ldots,N$, we have
 \begin{align}\label{eq:bond}
  &\xi_i=1,\quad a_i(t)=\varepsilon_i,\quad Z_i(t)=1-L_i(t),\quad K_i(t)=H_i(t);\\
  &\xi_{N+1}=0,\quad  a_{N+1}(t)=-\varepsilon_{N+1},\quad Z_{N+1}(t)=L_{N+1}(t),\quad K_{N+1}(t)=H_{N+1}(t).\nonumber
 \end{align}
As for the CDS portfolio, we also have that $\bar{N}=N$, and $\bar{\tau}_i=\tau_i$ for $i=1,\ldots,N+1$.

{\bf First-to-default claim.}\quad We have
\begin{align}\label{eq:first-to}
  &\xi_1=0,\quad a_1(t)=\varepsilon,\quad Z_1(t)=\sum_{i=1}^N L_i(t)H_i(t),\quad K_1(t)=1-\prod_{i=1}^N(1-H_i(t));\\
   &\xi_2=0,\quad a_2(t)=-\varepsilon_{N+1},\quad Z_2(t)=L_{N+1}(t),\quad K_2(t)=H_{N+1}(t).\nonumber
 \end{align}
Hence, in terms of the Definition~\ref{def:port-def}, we have that $\bar{N}=1$, $\bar{\tau}_1=\tau_1\wedge\cdots\wedge\tau_N$ and $\bar{\tau}_2=\tau_{N+1}$.
\end{example}

For $i=1,\ldots,\bar{N}+1$, {and $t\in[0,T]$,} let $S_i(t,T)$ be the {time $t$} price of the $i$-th defaultable claim $(\xi_i,a_i,Z_i,K_i)$ in the portfolio. By virtue of Proposition~\ref{lem:pre-default} and the expressions in Eq.~\eqref{eq:FalphaF12}, it follows that for $i=1,\ldots,\bar{N}+1$, on $\{\bar{\tau}_i>t\}$,
\begin{align}\label{eq:StT2i}
S_i(t,T)={\bf1}_{t\neq T}F_{i;(1,0,0)}(t,X(t),H(t))+F_{i;(0,1,1)}(t,X(t),H(t))-Z_i(t)K_i(t).
\end{align}
Above, the functions $F_{i;(1,0,0)}(t,x,z)$ and $F_{i;(0,1,1)}(t,x,z)$ are the unique bounded classical solutions to the following recursive system of backward Cauchy problems in which we set, respectively, $\alpha=(1,0,0)$ and $\alpha=(0,1,1)$: on $(t,x,z)\in[0,T)\times\R_+^{N+1}\times{\cal S}$,
\begin{align}\label{eq:Cahchypro1i}
&\left(\frac{\partial}{\partial t} + \A\right)F_{i;\alpha}(t,x,z)+\alpha_3(1-K_i(z))a_i(z)-\alpha_3\sum_{j=1}^{N+1}K_i(z)[Z_i(z^j)-Z_i(z)](1-z_j)x_j= 0
\end{align}
with terminal condition
\begin{align}\label{eq:terminal-conditioni}
F_{i;\alpha}(T,x,z)=\alpha_1\xi_i(z)(1-K_i(z))+\alpha_2Z_i(z)K_i(z),\qquad (x,z)\in \R_+^{N+1}\times{\cal S}.
\end{align}

{We next derive the analytical representation of the CVA process using the sequence of price processes generated by a defaultable claim portfolio.} We first define the exposure of the investor (assumed default-free) to the counterparty, ``$N+1$''. This represents the loss that the hedger would incur if his counterparty ``$N+1$'' would default {before or }at time $t$. It is given by
\begin{equation}
\varepsilon_{\bar{N}}(t,T):=\sum_{i=1}^{\bar{N}} b_i S_i(t,T),\label{eq:expos}
\end{equation}
where for $i=1,\ldots,\bar{N}$, the weight $b_i\in\R$
indicates the number of contracts referencing the entity $i$ purchased ($b_i > 0$) or sold ($b_i < 0$) by the investor. Therefore from Eq.~\eqref{eq:expos}, it holds that
\begin{align}\label{eq:expos2}
\varepsilon_{\bar{N}}(t,T)&=\sum_{i=1}^{\bar{N}} b_i S_i(t,T) {\bf1}_{\bar{\tau}_i \geq t}=\sum_{i=1}^{\bar{N}} b_i S_i(t\wedge\bar{\tau}_i,T) {\bf1}_{\bar{\tau}_i \geq t}\nonumber\\
&=\sum_{i=1}^{\bar{N}} b_i\big[(1-K_i(t))S_i(t,T)+Z_i(\bar{\tau}_i){\bf1}_{\bar{\tau}_i=t}\big].
\end{align}
Since $K_1(t),\ldots,K_{\bar{N}+1}(t)$ do not jump simultaneously (see Definition~\ref{def:port-def}), {and noticing that $K_{\bar{N}+1}(t)=H_{N+1}(t)$ is the default indicator process of the counterparty}, we obtain from \eqref{eq:expos2} that
\begin{align}\label{eq:eNtau}
\varepsilon_{\bar{N}}(\tauN,T) =\sum_{i=1}^{\bar{N}} b_i (1-K_i(\tauN))S_i(\tauN,T).
\end{align}
The CVA of the defaultable claim portfolio is given by
\begin{equation*}
{\rm CVA}_{\bar{N}}(t,T) =  \Ex\big[ L_{N+1}(\tauN) {\bf1}_{\{t<\tauN\leq T\}}\{\varepsilon_{\bar{N}}(\tauN,T)\}_+\big|\G_t\big],
\end{equation*}
where $x_+=x \vee 0$ for any real number $x\in\R$, i.e. the positive part of the real number $x$. Above, $L_{N+1}(t)=L_{N+1}(H(t))$ denotes the percentage loss rate incurred by the investor when counterparty ``$N+1$'' defaults on its obligations. This loss is paid at the default time $\tauN$. {As we are considering dynamic hedging of CVA, this may be seen as} a payment stream on the random interval $[0, T \wedge \tauN]$. More precisely, its payment stream $\Theta=(\Theta(t))_{t\in[0,T]}$ is given by
\begin{align}\label{eq:CVA-stream1def}
\left\{
  \begin{array}{ll}
   {\Theta}(t)= L_{N+1}(\tauN) {\bf1}_{\tauN\leq t}\{\varepsilon_{\bar{N}}(\tauN,T)\}_+,\ \ \ t\in[0,T),\\ \\
   {\Theta}(T)=0,\ \ t=T,
  \end{array}
\right.
\end{align}
where the exposure $\varepsilon_{\bar{N}}(\tauN,T)$ is given by Eq.~\eqref{eq:eNtau}.

The hedging instrument herein is chosen to be the gain process of the CDS contract referencing the counterparty. Next, we derive the dynamics of the gain process $Y_{N+1}=(Y_{N+1}(t))_{t\in[0,T]}$ of the CDS. Recall the representation of the CDS portfolio $(\xi_i,a_i,Z_i,K_i)_{i=1,\ldots,N+1}$ given by Eq.~\eqref{eq:CDS}. Let $Y_i(t):=\Ex[D_i(t)|\G_t]$ be the gain process of the $i$-th CDS for $i=1,\ldots,N+1$. For the CDS portfolio given by Eq.~\eqref{eq:CDS}, it follows that, for $i=1,\ldots,N+1$,
\begin{align*}
\sum_{j=1}^{N+1}K_i(z)[Z_i(z^j)-Z_i(z)](1-z_j)x_j = \sum_{j=1}^{N+1}z_i[L_i(z^j)-L_i(z)](1-z_j)x_j=\sum_{j\neq i}z_i[L_i(z^j)-L_i(z)](1-z_j)x_j.
\end{align*}
{Recall the system of Cauchy problems given by}~\eqref{eq:Cahchypro1i} and \eqref{eq:terminal-conditioni}. For $i=1,\ldots,N+1$, consider the following recursive system of backward Cauchy problems associated with the CDS portfolio: on $(t,x,z)\in[0,T)\times\R_+^{N+1}\times{\cal S}$,
\begin{align}\label{eq:Cahchypro1CDS}
&\left(\frac{\partial}{\partial t} + \A\right)F_{i}^{{\rm cds}}(t,x,z)-(1-z_i)\varepsilon_i-\sum_{j\neq i}z_i[L_i(z^j)-L_i(z)](1-z_j)x_j= 0
\end{align}
with terminal condition
\begin{align}\label{eq:terminal-conditionCDS}
F_{i}^{{\rm cds}}(T,x,z)=Z_i(z)K_i(z)=L_i(z)z_i,\qquad (x,z)\in \R_+^{N+1}\times{\cal S}.
\end{align}
Therefore Lemma~\ref{lem:gain} gives that
\begin{lemma}\label{lem:gainCDS}
Under the assumptions ({\bf A1}) and ({\bf A2}), for $i=1,\ldots,N+1$, the gain process $Y_i(t) = \Ex\left[D_i(T) |\G_t\right]$ of the $i$-th CDS, i.e. associated with the defaultable claim $(\xi_i,a_i,Z_i,K_i)$ specified in~\eqref{eq:CDS}, admits the following dynamics, for $t\in[0,T]$,
\begin{align}\label{eq:YiCDS}
dY_i(t) &= V_i^{{\rm cds}}(t,X(t),H(t))^{\top} \sigma(X(t))\D W(t)\\
&\quad+  \sum_{j=1}^{N+1}\big\{G_{ij}^{{\rm cds}}(t,X(t^-),H(t^-))-H_i(t^-)[L_i^j(t^-)-L_i(t^-)]\big\}\D M_j(t).\nonumber
\end{align}
Above, we recall that $L_i(t):=L_i(H(t))$ and $L_i^j(t):=L_i(H^j(t))$. For $j=1,\ldots,N+1$, the process $M_j=(M_j(t))_{t\in[0,T]}$ is the $\Gx$-default martingale given by Eq.~\eqref{eq:default-mart1C}. For $(t,x,z)\in[0,T]\times\R_+^{N+1}\times{\cal S}$,
\begin{align}\label{eq:VGCDS}
V_i^{\rm cds}(t,x,z)&:=D_xF_{i}^{{\rm cds}}(t,x,z),\qquad
G_{ij}^{\rm cds}(t,x,z):=F_{i}^{\rm cds}(t,x+w_j,z^j) - F_{i}^{\rm cds}(t,x,z),
\end{align}
for all $i,j=1,\ldots,N+1$.
\end{lemma}

\begin{remark}\label{rem:CDSs-dyna}Consider the special case that $Z_i(z)=L_i(z)=L_i$ for $i=1,\ldots,N+1$, i.e. they are constants and independent of the default state $z\in{\cal S}$. The Cauchy system \eqref{eq:Cahchypro1CDS} then reduces to, on $(t,x,z)\in[0,T)\times\R_+^{N+1}\times{\cal S}$,
\begin{align}\label{eq:Cahchypro1CDSLi}
&\left(\frac{\partial}{\partial t} + \A\right)F_{i}^{{\rm cds}}(t,x,z)-(1-z_i)\varepsilon_i= 0
\end{align}
with terminal condition $F_{i}^{{\rm cds}}(T,x,z)=L_iz_i$ for $(x,z)\in \R_+^{N+1}\times{\cal S}$. Correspondingly the gain process $Y_i(t) = \Ex\left[D_i(T) |\G_t\right]$ of the $i$-th CDS admits dynamics
\begin{align}\label{eq:YiLi}
dY_i(t) &= V_i^{{\rm cds}}(t,X(t),H(t))^{\top} \sigma(X(t))\D W(t)+\sum_{j=1}^{N+1}G_{ij}^{{\rm cds}}(t,X(t^-),H(t^-))\D M_j(t).
\end{align}
\end{remark}

\begin{lemma}\label{lem:stopedTheta}
The stopped payment stream related associated with the CVA of a defaultable claim portfolio $(\xi_i,a_i,Z_i,K_i)_{i=1,\ldots,\bar{N}+1}$ before maturity admits the analytical representation:
\begin{align}\label{eq:decom1}
&\Theta(\tauN\wedge T)=\int_0^T{\bf1}_{s<T}L_{N+1}^{N+1}(s^-)\\
&\qquad\times\Bigg\{\sum_{i=1}^{\bar{N}} b_i \big(1-K_i^{N+1}(s^-)\big)F_{i}\big(s,X(s^-)+w_{N+1},H^{N+1}(s^-)\big)\Bigg\}_+dH_{N+1}(s),\nonumber
\end{align}
where $F_i(t,x,z)$ is the unique bounded classical solution to the recursive system \eqref{eq:Cahchypro1i} of the backward Cauchy problems in which we set $\alpha=(1,1,1)$,  i.e., $F_i(t,x,z):=F_{i;(1,1,1)}(t,x,z)$. 
We also recall the notations introduced in \eqref{eq:trasferdefault}.
\end{lemma}

\noindent{\it Proof.}\quad We notice from Eq.~\eqref{eq:CVA-stream1def} that for $t\in[0,T]$,
\begin{align*}
{\Theta}(t)=\Theta(t){\bf1}_{t<T}= L_{N+1}(\tauN) {\bf1}_{\tauN\leq t}\{\varepsilon_{\bar{N}}(\tauN,T)\}_+{\bf1}_{t<T}.
\end{align*}
Then we have
\begin{align*}
\Theta(\tauN\wedge T)&=\Theta(\tauN){\bf1}_{\tauN \leq T}\nonumber\\
&=L_{N+1}(\tauN) {\bf1}_{\tauN\leq \tauN}\{\varepsilon_{\bar{N}}(\tauN,T)\}_+{\bf1}_{\tauN<T}{\bf1}_{\tauN\leq T}\nonumber\\
 &=L_{N+1}(\tauN)\{\varepsilon_{\bar{N}}(\tauN,T)\}_+{\bf1}_{\tauN<T}\nonumber\\
&=L_{N+1}(\tauN)\{\varepsilon_{\bar{N}}(\tauN,T)\}_+{\bf1}_{\tauN\leq T},
\end{align*}
where we used the fact that $S_i(T,T)=0$ for all $i=1,\ldots,\bar{N}$ using the price representation \eqref{eq:StT2i}. Hence, it holds that $\varepsilon_{\bar{N}}(T,T)=0$. It thus follows from the price representation \eqref{eq:StT2i} that
\begin{align}\label{eq:EQ2}
\Theta(\tauN\wedge T)&=\int_0^TL_{N+1}(s)\{\varepsilon_{\bar{N}}(s,T)\}_+dH_{N+1}(s)\nonumber\\
&=\int_0^T{\bf1}_{s<T}L_{N+1}(s)\left\{\sum_{i=1}^{\bar{N}} b_i (1-K_i(s))S_i(s,T)\right\}_+dH_{N+1}(s)\nonumber\\
&=\int_0^T{\bf1}_{s<T}L_{N+1}(s)\Bigg\{\sum_{i=1}^{\bar{N}} b_i (1-K_i(s))\big[{\bf1}_{s\neq T}F_{i;(1,0,0)}(s,X(s),H(s))\nonumber\\
&\qquad+F_{i;(0,1,1)}(t,X(s),H(s))-Z_i(H(s))K_i(H(s))]\bigg\}_+dH_{N+1}(s)\nonumber\\
&=\int_0^T{\bf1}_{s<T}L_{N+1}(s)\Bigg\{\sum_{i=1}^{\bar{N}} b_i (1-K_i(s))\big[F_{i;(1,0,0)}(s,X(s),H(s))\nonumber\\
&\qquad+F_{i;(0,1,1)}(t,X(s),H(s))-Z_i(H(s))K_i(H(s))]\bigg\}_+dH_{N+1}(s).
\end{align}
Notice that $F_{i;(1,0,0)}(t,x,z)+F_{i;(0,1,1)}(t,x,z)=F_{i;(1,1,1)}(t,x,z)$. Thus it holds that
\begin{align*}
&\Theta(\tauN\wedge T)\nonumber\\
&=\int_0^T{\bf1}_{s<T}L_{N+1}(s)\left\{\sum_{i=1}^{\bar{N}} b_i (1-K_i(s))\left[F_{i;(1,1,1)}(s,X(s),H(s))-Z_i(H(s))K_i(s)\right]\right\}_+dH_{N+1}(s)\nonumber\\
&=\int_0^T{\bf1}_{s<T}L_{N+1}(H^{N+1}(s^-))\Bigg\{\sum_{i=1}^{\bar{N}} b_i (1-K_i(H^{N+1}(s^-)))\\
&\qquad\times\left[F_{i;(1,1,1)}(s,X(s^-)+w_{N+1},H^{N+1}(s^-))-Z_i(H^{N+1}(s^-))K_i(H^{N+1}(s^-))\right]\Bigg\}_+dH_{N+1}(s).\nonumber
\end{align*}
This yields the representation~\eqref{eq:decom1} using that $F_i(t,x,z)=F_{i;(1,1,1)}(t,x,z)$ and $(1-K_i(z))K_i(z)=0$. Hence, the proof of the lemma is completed. \hfill$\Box$

\section{Risk-Minimizing Hedging for CVA}\label{sec:hedge}

This section studies dynamic hedging of the CVA for a defaultable claim portfolio of the form given in Definition~\ref{def:port-def}. The hedging instrument used by the investor is the CDS written on the investor's counterparty ``$N+1$'' and a riskless asset. In our incomplete market model,
the existence of a self-financing strategy that perfectly replicates the CVA is not guaranteed. We thus choose to implement an optimal hedging strategy that perfectly replicates the CVA claim, but with a small cost, such that it remains self-financing on average.

Recall that $\Theta=(\Theta(t))_{t\in[0,T]}$ is the CVA payment stream associated with the defaultable claims portfolio given by $(\xi_i,a_i,Z_i,K_i)_{i=1,\ldots,\bar{N}+1}$, and given in Eq.~\eqref{eq:CVA-stream1def}. Hedging is performed until the CVA payoff is triggered. Hence, we work with hedging strategies only up to $T\wedge \tauN$, i.e. the minimum between the maturity of the CVA claim and the default time of the investor's counterparty. As in \cite{FreySchimdt}  and \cite{Freyback}, we use the gain process as hedging instrument: in our framework, this is given
by the process  $Y_{N+1}=(Y_{N+1}(t))_{t\in[0,T]}$ considered in Lemma~\ref{lem:gainCDS} under the choice $i=N+1$. That is, the dynamics of  $Y_{N+1}=(Y_{N+1}(t))_{t\in[0,T]}$ is given by
\begin{align}\label{eq:YN+1}
dY_{N+1}(t) &= V_{N+1}^{{\rm cds}}(t,X(t),H(t))^{\top} \sigma(X(t))\D W(t)\\
&\quad+  \sum_{j=1}^{N+1}\big\{G_{N+1,j}^{{\rm cds}}(t,X(t^-),H(t^-))-H_{N+1}(t^-)[L_{N+1}^{j}(t^-)-L_{N+1}(t^-)]\big\}\D M_j(t).\nonumber
\end{align}
We recall that $V_{N+1}^{{\rm cds}}(t,x,z)$ and $G_{N+1,j}^{{\rm cds}}(t,x,z)$ are given by \eqref{eq:VGCDS}, choosing $i=N+1$.

\begin{definition}\label{def:Psi}
Let $\Psi$ be the space of all $\Gx$-predictable processes $\theta=(\theta(t))_{t \in [0, T\wedge \tauN]}$ such that
\begin{align*}
\Ex\left[\int_0^{T\wedge\tau_{N+1}}\theta^2(t) d\left<Y_{N+1},Y_{N+1}\right>(t) \right ] < \infty.
\end{align*}
An admissible strategy is a bidimensional process $\varphi =(\theta, \eta)$  where  $\theta \in \Psi$ and $\eta$ is a real-valued $\Gx$-adapted process such that the associated value process
$V^{\varphi}(t) := \theta(t) Y_{N+1}(t)  + \eta(t)$ is right-continuous and square integrable over $[0, T\wedge \tauN]$.
\end{definition}
Above, $\theta(t)$ denotes the number of shares of the gain process of the risky CDS contract referencing the counterparty held at time $t$, while $\eta(t)$ is the amount invested in the riskless asset at time $t$. Following \cite{Schweizer08} who investigates the case of payment streams over a deterministic time horizon, and \cite{BiaCre} who allow for a random delivery date which can be seen as a payment stream over a random time horizon, we assign a cost process to each admissible strategy:

\begin{definition}
The cost process $C^{\varphi}$ of an admissible strategy $\varphi=(\theta, \eta)$ is given by
\begin{equation}\label{eq:cost}
C^{\varphi}(t) := \Theta(t )+ V^{\varphi}(t)  - \int_0^t \theta(u) dY_{N+1}(u) , \quad t \in [0, T\wedge \tauN],\end{equation}
where $\Theta(t)$ is defined in \eqref{eq:CVA-stream1def}. An admissible strategy $\varphi$ is called mean-self-financing if its cost process $C^{\varphi}$ is a martingale. The risk process of $\varphi$, that is the conditional variance of the hedging error, is given by
\begin{align*}
R^{\varphi}(t) := \Ex \left[ \big(C^{\varphi}(T\wedge\tauN) -  C^{\varphi}(t)\big)^2 \big| \G_t \right], \quad t \in\Ttau.
\end{align*}
\end{definition}
It is well known in the literature that a natural extension of the risk-minimization approach to payment streams requires to look for admissible strategies with the $0$-achieving property, that is such that
$V^{\varphi}(\tauN  \wedge T)=0.$
\begin{definition}
Let $\Theta$ be the payment stream given in Eq.~\eqref{eq:CVA-stream1def}. We say that an admissible strategy  $\varphi^*$ is risk minimizing for $\Theta$ if the following conditions hold:
\begin{itemize}
\item[(i)]$\varphi^*$ is $0$-achieving, that is $V^{\varphi^*}(\tauN \wedge T)=0$;
\item[(ii)] $\varphi^*$ minimizes the risk process $R^{\varphi}$ over the class of admissible strategies.
\end{itemize}
\end{definition}
Let us consider the CVA payment stream $\Theta(t)$  given in Eq.~\eqref{eq:CVA-stream1def}. Notice that $\Theta(t)$ is square integrable for all $t\in[0,T]$ since the price representation $S_i(t,T)$ given by Eq.~\eqref{eq:StT2i} is bounded for all $i=1,\ldots,\bar{N}$ using Proposition~\ref{lem:sol-Fi2}. Hence we can write the GKW decomposition  of $ \Theta(T\wedge\tauN )$ with respect to the martingale $Y_{N+1}$. This is given by
\begin{equation}\label{eq:GKW}
 \Theta(T\wedge\tauN ) =  \Ex \left[  \Theta(T\wedge\tauN ) \right]  +\int_0^{T\wedge\tauN} \theta^{GKW}(u) dY_{N+1}(u) + A(T \wedge\tauN ),
\end{equation}
where $\theta^{GKW}$ is a $\Gx$-predictable, integrable process with respect to $Y_{N+1}$, and $A$ is a martingale null at time zero, strongly orthogonal to $Y_{N+1}$. Define the process  $V$ by setting
\begin{equation}\label{eq:V}
V(t) :=\Ex \left[ \Theta(T\wedge\tauN ) | \G_t \right],\quad \ t \in \Ttau.
\end{equation}
By conditioning on $\G_t$ in Eq.~\eqref{eq:GKW}, we obtain
\begin{align}\label{eq:GKWbis}
V(t)=\Ex \left[  \Theta(T\wedge\tauN ) \right] +\int_0^t \theta^{GKW}(u) dY_{N+1}(u) + A(t),  \quad t \in  \Ttau.
\end{align}
The following proposition, whose proof is postponed to the Appendix, establishes the connection between the GKW decomposition  of $ \Theta(T\wedge\tauN)$ given in Eq.~\eqref{eq:GKW} and the risk-minimizing strategy for the payment stream $\Theta(t)$ associated with the CVA contract. Such a result extends Theorem 2.4 in \cite{guided} to the case of payment streams with a random delivery date. For the case of local-risk minimization, the proof can be found in \cite{Schweizer08}.
\begin{prop}\label{thm:strategy}
The payment stream $\Theta$ given by Eq.~\eqref{eq:CVA-stream1def} admits a unique risk-minimizing strategy $\varphi^* = (\theta^*, \eta^*)$, where for $t\in\Ttau$,
\begin{align}\label{eq:thetaGKW}
\theta^*(t) = \theta^{GKW}(t),\quad\text{and }\  \eta^*(t)= V(t) -  \Theta(t)+ \theta^{GKW}(t)Y_{N+1}(t).
\end{align}
The optimal value process and the minimal cost are given by
\begin{align}\label{eq:thetaGKWV}
V^{\varphi^*}(t) = V(t) - \Theta(t),\ \text{ and }\ C^{\varphi^*}(t) = \Ex \left[  \Theta(T\wedge\tauN) \right] + A(t).
\end{align}
Moreover, the strategy $\theta^{GKW}$ admits the following representation, for $t\in\Ttau$,
\begin{align}\label{eq:thetaGKW2}
\theta^{GKW}(t)={ d \left< V, Y_{N+1}\right>(t) \over d\left<Y_{N+1},Y_{N+1}\right>(t)}.
\end{align}
\end{prop}

Our next goal is to provide a more explicit representation for the process $\theta^{GKW}=(\theta^{GKW}(t))_{t\in\Ttau}$ given in Eq.~\eqref{eq:thetaGKW2}. We start providing the martingale decomposition of the process $V$ defined by Eq.~\eqref{eq:V}, which will be given in {Proposition} \ref{thm:mart-V} below.  Toward this goal, we consider existence and uniqueness of classical solutions to a recursive system of Cauchy problems, which will play an important role for the representation of the process $\theta^{GKW}$ given in~\eqref{eq:thetaGKW2}. We also study the boundedness of these solutions, which serves to guarantee that the risk minimizing strategy associated with the process $\theta^{GKW}$ belongs to the space $\Psi$ given in Definition~\ref{def:Psi}. For any $z\in{\cal S}$, on $(t,x)\in[0,T)\times\R_+^{N+1}$,
\begin{align}\label{eq:F-K-g0}
0&=\left(\frac{\partial}{\partial t}+{\cal A}\right)g(t,x,z)\nonumber\\
&\quad+L_{N+1}(z^{N+1})\left\{\sum_{i=1}^{\bar{N}} b_i(1-K_i(z^{N+1}))F_{i}(t,x+w_{N+1},z^{N+1})\right\}_+(1-z_{N+1})x_{N+1}
\end{align}
with terminal condition $g(T,x,z)=0$ for all $(x,z)\in\R_+^{N+1}\times{\cal S}$. Above, the operator ${\cal A}$ is defined by \eqref{eq:A}, and $F_i(t,x,z)$ is the unique bounded classical solution to the system \eqref{eq:Cahchypro1i} in which we take $\alpha=(1,1,1)$. Rewrite Eq.~\eqref{eq:F-K-g0} in a more convenient form:
\begin{align}\label{eq:F-K-g2}
0&=\left(\frac{\partial}{\partial t}+\tilde{\cal A}\right)g(t,x,z)+\sum_{j=1}^{N+1} \big[g(t,x+w_j,z^j) - g(t,x,z)\big](1-z_j)x_j\nonumber\\
&\quad+L_{N+1}(z^{N+1})\left\{\sum_{i=1}^{\bar{N}} b_i(1-K_i(z^{N+1}))F_{i}(t,x+w_{N+1},z^{N+1})\right\}_+(1-z_{N+1})x_{N+1},
\end{align}
where $\tilde{\cal A}$ is given by \eqref{eq:tildeA}. Similarly to the recursive system of Cauchy problems \eqref{eq:Cahchypro1}, we can study the solvability of Eq.~\eqref{eq:F-K-g2} recursively through the default states $z=0^{j_1,\ldots,j_l}$ for $l=0,1,\ldots,N+1$. We also define $g^{(l)}(t,x)$ and $g^{(l+1),i}(t,x)$ by \eqref{eq:flfl1i} with $f$ replaced by $g$. It may be easily seen that when $l=N+1$, Eq.~\eqref{eq:F-K-g2} simplifies to
\begin{align*}
\left(\frac{\partial}{\partial t}+\tilde{\cal A}\right)g^{(N+1)}(t,x)=0
\end{align*}
with terminal condition $g^{(N+1)}(T,x)=0$ for all $x\in\R_+^{N+1}$. It can be immediately verified that this admits the solution $g^{(N+1)}(t,x)=0$ for all $(t,x)\in[0,T]\times\R_+^{N+1}$. In the more general case that $z=0^{j_1,\ldots,j_l}$ where $l=0,1,\ldots,N$, we need to deal with the following cauchy problem defined on the unbounded domain: on $(t,x)\in[0,T)\times\R_+^{N+1}$,
\begin{align}\label{eq:F-K-g2z=m}
0&=\left(\frac{\partial}{\partial t}+\tilde{\cal A}\right)g^{(l)}(t,x)-\left(\sum_{j\notin\{j_1,\ldots,j_l\}}x_j\right)g^{(l)}(t,x)+\sum_{j\notin\{j_1,\ldots,j_l\}}g^{(l+1),j}(t,x+w_j)x_j\nonumber\\
&\quad+L_{N+1}^{(l+1),N+1}\left\{\sum_{i=1}^{\bar{N}} b_i(1-K_i^{(l+1),N+1})F_{i}(t,x+w_{N+1},0^{j_1,\ldots,j_l,N+1})\right\}_+x_{N+1}{\bf1}_{j_1,\ldots,j_l\neq N+1}
\end{align}
with terminal condition $g^{(l)}(T,x)=0$ for all $x\in\R_+^{N+1}$. The function $g^{(l+1),j}(t,x)$ is the unique classical solution of the Cauchy system~\eqref{eq:F-K-g2} when the default state $z=0^{j_1,\ldots,j_l,j}$, for $j\notin\{j_1,\ldots,j_l\}$. Recall the notation $L_{N+1}^{(l+1),N+1}=L_{N+1}(0^{j_1,\ldots,j_l,N+1})$ and $K_i^{(l+1),N+1}=K_i(0^{j_1,\ldots,j_l,N+1})$ for $j_1,\ldots,j_l\neq N+1$.

Existence and uniqueness of (nonnegative) bounded classical solutions to the Cauchy problem~\eqref{eq:F-K-g2z=m} can be proven inductively as stated in the following theorem. {The proof is reported in the Appendix.}
\begin{theorem}\label{thm:sol}
Let assumptions ({\bf A1}) and ({\bf A2}) hold.
Assume that for $j\notin\{j_1,\ldots,j_l\}$, $l=0,1,\ldots,N$, the Cauchy system~\eqref{eq:F-K-g2} admits a unique (nonnegative) bounded classical solution $g^{(l+1),j}(t,x)$ when $z=0^{j_1,\ldots,j_l,j}$. Then the Cauchy system~\eqref{eq:F-K-g2} also admits a unique (nonnegative) bounded classical solution $g^{(l)}(t,x)$ when $z=0^{j_1,\ldots,j_l}$ (i.e., the Cauchy problem \eqref{eq:F-K-g2z=m} above admits a unique (nonnegative) bounded classical solution).
\end{theorem}
Using Theorem~\ref{thm:sol}, we obtain the following martingale decomposition of the process $V$ defined by Eq.~\eqref{eq:V}:
\begin{prop}\label{thm:mart-V}
Let assumptions ({\bf A1}) and ({\bf A2}) hold. The process $V$ defined by Eq.~\eqref{eq:V} admits the martingale decomposition given by, for $t\in[0,T]$,
\begin{align}\label{eq:sde-V}
V(t)&=V(0)+\int_0^tD_{x}g(s,X(s),H(s))^{\top}\sigma(X(s))\D W(s)\nonumber\\
&\quad+\int_0^t{\bf1}_{s<T}L_{N+1}^{N+1}(s^-)\\
&\qquad\quad\times\left\{\sum_{i=1}^{\bar{N}} b_i (1-K_i^{N+1}(s^-))F_{i}(s,X(s^-)+w_{N+1},H^{N+1}(s^-))\right\}_+dM_{N+1}(s)\nonumber\\
&\quad+\sum_{j=1}^{N+1}\int_0^t\left[g(t,{X}(s^-)+w_j,H^j(s^-))-g(t,{X}(s^-),H(s^-))\right]\D M_j(s).\nonumber
\end{align}
For $j=1,\ldots,N+1$, we recall that the process $M_j=(M_j(t))_{t\in[0,T]}$ is the $\Gx$-default martingale given by Eq.~\eqref{eq:default-mart1C}, and for $(t,x,z)\in[0,T]\times\R_+^{N+1}\times{\cal S}$, the function $g(t,x,z)$ is the unique nonnegative bounded classical solution to the recursive system of Cauchy problems~\eqref{eq:F-K-g0}. The function $F_i(t,x,z)$ is the unique bounded classical solution to the recursive system \eqref{eq:Cahchypro1i} in which we set $\alpha=(1,1,1)$, and $D_xg(t,x,z)$ is a column vector denoting the gradient of $g(t,x,z)$ w.r.t. $x$.
\end{prop}

\noindent{\it Proof.}\quad We first have that $\Theta( T)= \Theta(\tauN\wedge T)$ and $\Theta( T)=0$ on $\{ \tau_{N+1} > T\}$.  By virtue of Lemma~\ref{lem:stopedTheta}, for $t\in[0, T\wedge\tau_{N+1}]$, it holds that
\begin{align*}
&V(t) =\Ex\left[\Theta(\tauN\wedge T) | \G_t\right]=\Ex\left[\int_0^T{\bf1}_{s<T}\Upsilon_sdH_{N+1}(s)\right| \G_t\Bigg],
\end{align*}
where the $\Gx$-predictable process
\begin{align*}
\Upsilon_s:=L_{N+1}(H^{N+1}(s^-))\left\{\sum_{i=1}^{\bar{N}} b_i (1-K_i(H^{N+1}(s^-)))F_{i}(s,X(s^-)+w_{N+1},H^{N+1}(s^-))\right\}_+.
\end{align*}
Then we have
\begin{align}\label{eq:Vt}
V(t) &=\Ex\Bigg[\int_0^T{\bf1}_{s<T}  \Upsilon_s dM_{N+1}(s)\bigg| \G_t\Bigg]
+\Ex\Bigg[\int_0^T  \Upsilon_s (1-H_{N+1}(s^-))X_{N+1}(s^-)ds\bigg| \G_t\Bigg] \nonumber \\
&=\int_0^t {\bf1}_{s<T}  \Upsilon_s dM_{N+1}(s)
+\int_0^t  \Upsilon_s (1-H_{N+1}(s))X_{N+1}(s)ds +V_2(t).
\end{align}
Above, the process $V_2=(V_2(t))_{t\in[0,T]}$ is defined by, for $t\in[0,T]$,
\begin{align*}
V_2(t)&:=\Ex\left[\int_t^T\Upsilon_s(1-H_{N+1}(s))X_{N+1}(s)ds\bigg| \G_t\right].
\end{align*}
We next provide an explicit characterization of the above process $V_2$. Let $(t,x,z)\in[0,T]\times\R_+^{N+1}\times{\cal S}$ and define
\begin{align}\label{eq:v2}
g(t,x,z)&:=\Ex_{t,x,z}\left[\int_t^T\Upsilon_s(1-H_{N+1}(s))X_{N+1}(s)ds\right].
\end{align}
Because $(X,H)$ is a $\Gx$-Markov process, we have that $V_2(t)=g(t,X(t),H(t))$ for $t\in[0,T]$. Moreover, using the Feymann-Kac's formula, it follows that $g(t,x,z)$ satisfies the Cauchy problem~\eqref{eq:F-K-g0}, i.e., on $(t,x,z)\in[0,T)\times\R_+^{N+1}\times{\cal S}$,
\begin{align*}
0&=\left(\frac{\partial}{\partial t}+{\cal A}\right)g(t,x,z)\nonumber\\
&\quad +L_{N+1}(z^{N+1})\left\{\sum_{i=1}^{\bar{N}} b_i(1-K_i(z^{N+1}))F_{i}(t,x+w_{N+1},z^{N+1})\right\}_+(1-z_{N+1})x_{N+1}
\end{align*}
with terminal condition $g(T,x,z)=0$ for all $(x,z)\in\R_+^{N+1}\times{\cal S}$. Thanks to Theorem~\ref{thm:sol}, we can apply It\^o's formula and obtain
\begin{align*}
g(t,X(t),H(t))&=g(0,X(0),H(0))+\int_0^t\left(\frac{\partial}{\partial s}+{\cal A}\right)g(s,X(s),H(s))ds\nonumber\\
 &\quad+\int_0^tD_{x}g(s,X(s),H(s))^{\top}\sigma(X(s))\D W(s)\\
&\quad+\sum_{j=1}^{N+1}\int_0^t\left[g(s,{X}(s^-)+w_j,H^j(s^-))-g(s,{X}(s^-),H(s^-))\right]\D M_j(s).\nonumber
\end{align*}
Then the Cauchy problem~\eqref{eq:F-K-g0} implies that
\begin{align}\label{eq:dg}
dg(t,X(t),H(t))&=-L_{N+1}(H^{N+1}(t))\left\{\sum_{i=1}^{\bar{N}}b_i(1-K_i(H^{N+1}(t)))F_{i}(t,X(t)+w_{N+1},H^{N+1}(t))\right\}_+\nonumber\\
&\quad\quad\times(1-H_{N+1}(t))X_{N+1}(t)dt+D_{x}g(t,X(t),H(t))^{\top}\sigma(X(t))\D W(t)\nonumber\\
&\quad+\sum_{j=1}^{N+1}\left[g(t,{X}(t^-)+w_j,H^j(t^-))-g(t,{X}(t^-),H(t^-))\right]\D M_j(t).
\end{align}
Applying the decomposition \eqref{eq:Vt}, we obtain the martingale representation of $V(t)$ given by Eq.~\eqref{eq:sde-V}. \hfill$\Box$\\

We are now ready to give the characterize the risk-minimizing strategy for CVA.
\begin{theorem}\label{thm:thetaGWK}
Let assumptions ({\bf A1}) and ({\bf A2}) hold. The unique risk-minimizing strategy $\theta^{GKW} \in\Psi$ associated with the investment in the risky CDS contract referencing the counterparty ``$N+1$'' (see also Proposition~\ref{thm:strategy}) is given by
\begin{align}\label{eq:thetaGKW200}
\theta^{GKW}(t)&=\sum_{i=1}^3\frac{U_i(t,X(t^-),H(t^-))}{\Phi(t,X(t^-),H(t^-))},\qquad t\in\Ttau.
\end{align}
Above, for $(t,x,z)\in[0,T]\times\R_+^{N+1}\times{\cal S}$, the functions
\begin{align}\label{eq:coeffs}
U_1(t,x,z)&:=\big<D_{x}g(t,x,z)^{\top}\sigma(x),V_{N+1}^{\rm cds}(t,x,z)^{\top}\sigma(x)\big>;\nonumber\\
U_2(t,x,z)&:=L_{N+1}(z^{N+1})\left\{\sum_{i=1}^{\bar{N}} b_i(1-K_i(z^{N+1}))F_i(t,x+w_{N+1},z^{N+1})\right\}_+ G_{N+1,N+1}^{\rm cds}(t,x,z)x_{N+1};\nonumber\\
U_3(t,x,z)&:=\sum_{j=1}^{N}\left[g(t,x+w_j,z^j)-g(t,x,z)\right]\big\{G_{N+1,j}^{\rm cds}(t,x,z)-z_{N+1}[L_{N+1}(z^j)-L_{N+1}(z)]\big\}x_j(1-z_j)\nonumber\\
&\quad+\left[g(t,x+w_{N+1},z^{N+1})-g(t,x,z)\right]G_{N+1,N+1}^{\rm cds}(t,x,z)x_{N+1}.
\end{align}
The function
\begin{align}\label{eq:Phi}
\Phi(t,x,z)&:=\big|V_{N+1}^{\rm cds}(t,x,z)^{\top} \sigma(x)\big|^2\nonumber\\
&\quad+\sum_{j=1}^{N+1}\big|G_{N+1,j}^{{\rm cds}}(t,x,z)-z_{N+1}[L_{N+1}(z^j)-L_{N+1}(z)]\big|^2x_j(1-z_j).
\end{align}
For $(t,x,z)\in[0,T]\times\R_+^{N+1}\times{\cal S}$, the functions $V^{\rm cds}_{N+1}(t,x,z)$ and $G^{\rm cds}_{N+1,j}(t,x,z)$, $j=1,\ldots,N+1$, are given in \eqref{eq:VGCDS}. Recall that $g(t,x,z)$ is the unique nonnegative bounded classical solution to the recursive system of Cauchy problems~\eqref{eq:F-K-g0}. The function $F_i(t,x,z)$ is the unique bounded classical solution to the recursive system of backward Cauchy problems \eqref{eq:Cahchypro1i} in which we set $\alpha=(1,1,1)$.
\end{theorem}

\noindent{\it Proof.}
 \quad Recall that the dynamics of the gain process $Y_{N+1}=(Y_{N+1}(t))_{t\in[0,T]}$  of the CDS contract referencing the counterparty is given by Eq.~\eqref{eq:YN+1}, i.e.,
\begin{align*}
dY_{N+1}(t) &= V_{N+1}^{{\rm cds}}(t,X(t),H(t))^{\top} \sigma(X(t))\D W(t)\\
&\quad+  \sum_{j=1}^{N+1}\big\{G_{N+1,j}^{{\rm cds}}(t,X(t^-),H(t^-))-H_{N+1}(t^-)[L_{N+1}^{j}(t^-)-L_{N+1}(t^-)]\big\}\D M_j(t).\nonumber
\end{align*}
Then it holds that
\begin{align*}
d\left<Y_{N+1},Y_{N+1}\right>(t)=\Phi(t,X(t^-),H(t^-))dt.
\end{align*}
Here, for $(t,x,z)\in[0,T]\times\R_+^{N+1}\times{\cal S}$, the function $\Phi(t,x,z)$ is given by Eq.~\eqref{eq:Phi}. On the other hand, from  Theorem~\ref{thm:mart-V}, it follows that
\begin{align*}
d\left<V,Y_{N+1}\right>(t)&=U_1(t,X(t^-),H(t^-))dt+U_2(t,X(t^-),H(t^-))dt\nonumber\\
&\quad+U_3(t,X(t^-),H(t^-))dt,
\end{align*}
where, for $(t,x,z)\in[0,T]\times\R_+^{N+1}\times{\cal S}$, the functions $U_i(t,x,z)$, $i=1,2,3$, are given by Eq.~\eqref{eq:coeffs}.
Then from Eq.~\eqref{eq:thetaGKW2} in Proposition~\ref{thm:strategy}, it follows that for $t\in\Ttau$,
\begin{align*}
\theta^{GKW}(t)&={ d \left< V, Y_{N+1}\right>(t) \over d\left<Y_{N+1},Y_{N+1}\right>(t)}=\sum_{i=1}^3\frac{U_i(t,X(t^-),H(t^-))}{\Phi(t,X(t^-),H(t^-))}.
\end{align*}
This gives Eq.~\eqref{eq:thetaGKW200}.

We next verify $\theta^{GKW}\in{\Psi}$, i.e., $\theta^{GKW}$ is square integrable w.r.t. $d\left<Y_{N+1},Y_{N+1}\right>(t)$.
Below, we use $C$ to denote a generic positive constant, which may be different from line to line. We first notice that there exists a constant $C>0$ such that
\begin{align}\label{eq:thetine}
\Ex\left[\int_0^{T\wedge\tau_{N+1}}\left|\theta^{GKW}(t)\right|^2d\left<Y_{N+1},Y_{N+1}\right>(t)\right]\leq C\sum_{i=1}^3\Ex\left[\int_0^{T\wedge\tau_{N+1}}\frac{|U_i(t,X(t^-),H(t^-))|^2}{\Phi(t,X(t^-),H(t^-))}dt\right].
\end{align}
First, using H\"older's inequality, it holds that on $\Ttau$,
\begin{align*}
\frac{|U_1(t,X(t^-),H(t^-))|^2}{\Phi(t,X(t^-),H(t^-))}&\leq \frac{\left|V_{N+1}^{\rm cds}(t,X(t^-),H(t^-))^{\top} \sigma(X(t^-))\right|^2\left|D_xg(t,X(t^-),H(t^-))^{\top} \sigma(X(t^-))\right|^2}{\left|V_{N+1}^{\rm cds}(t,X(t^-),H(t^-))^{\top} \sigma(X(t^-))\right|^2}\nonumber\\
&=\left|D_xg(t,X(t^-),H(t^-))^{\top} \sigma(X(t^-))\right|^2.
\end{align*}
On the other hand, in terms of the expression~\eqref{eq:Phi} and the fact that $z_{N+1}(1-z_{N+1})=0$ for all $z_{N+1}\in\{0,1\}$, it follows that
\begin{align*}
\Phi(t,x,z)&\geq\sum_{j=1}^{N+1}\big|G_{N+1,j}^{{\rm cds}}(t,x,z)-z_{N+1}[L_{N+1}(z^j)-L_{N+1}(z)]\big|^2x_j(1-z_j)\nonumber\\
&\geq { \big|G_{N+1,N+1}^{{\rm cds}}(t,x,z)\big|^2}x_{N+1}(1-z_{N+1}).
\end{align*}
Then applying Proposition~\eqref{lem:sol-Fi2}, it holds that on $\Ttau$, there exists a constant $C>0$ such that
\begin{align*}
\frac{|U_2(t,X(t^-),H(t^-))|^2}{\Phi(t,X(t^-),H(t^-))}&\leq C\frac{\big| G_{N+1,N+1}^{\rm cds}(t,X(t^-),H(t^-))\big|^2 X^2_{N+1}(t^-)} {\big|G_{N+1,N+1}^{\rm cds}(t,X(t^-),H(t^-))\big|^2 X_{N+1}(t^-)}\leq CX_{N+1}(t^-).
\end{align*}
From Eq.~\eqref{eq:coeffs}, and noticing again that $z_{N+1}(1-z_{N+1})=0$ for $z_{N+1}\in\{0,1\}$, it follows that
\begin{align*}
U_3(t,x,z)&=\sum_{j=1}^{N+1}\left[g(t,x+w_j,z^j)-g(t,x,z)\right]\nonumber\\
&\quad\times\big\{G_{N+1,j}^{\rm cds}(t,x,z)-z_{N+1}[L_{N+1}(z^j)-L_{N+1}(z)]\big\}x_j(1-z_j).
\end{align*}
Using Theorem~\ref{thm:sol} and the Cauchy's inequality, it follows that on $\Ttau$ there exists a constant $C>0$ such that
\begin{align*}
\frac{|U_3(t,x,z)|^2}{\Phi(t,x,z)}&\leq C\frac{\big|\sum_{j=1}^{N+1}\{G_{N+1,j}^{\rm cds}(t,x,z)-z_{N+1}[L_{N+1}(z^j)-L_{N+1}(z)]\}x_j(1-z_j)\big|^2}
{\sum_{j=1}^{N+1}\big|G^{\rm cds}_{N+1,j}(t,x,z)-z_{N+1}[L_{N+1}(z^j)-L_{N+1}(z)]\big|^2x_j(1-z_j)}\nonumber\\
&\leq C\left(\sum_{j=1}^{N+1}x_j\right).
\end{align*}
This yields that
\begin{align*}
\frac{|U_3(t,X(t^-),H(t^-))|^2}{\Phi(t,X(t^-),H(t^-))}\leq C\left(\sum_{j=1}^{N+1}X_j(t^-)\right).
\end{align*}
Using the estimate \eqref{eq:thetine}, we deduce the existence of a constant $C>0$ such that
\begin{align}\label{eq:siest-2}
\Ex\left[\int_0^{T\wedge\tau_{N+1}}\left|\theta^{GKW}(t)\right|^2d\left<Y_{N+1},Y_{N+1}\right>(t)\right]&\leq C + C\Ex\left[\int_0^{T}\left|D_xg(t,X(t),H(t))^{\top} \sigma(X(t))\right|^2dt\right]\nonumber\\
&\quad+C\left\{\sum_{j=1}^{N+1}\Ex\left[\int_0^TX_j(t)dt\right]\right\}.
\end{align}

We next estimate the second term on the r.h.s. of the above inequality \eqref{eq:siest-2}. It follows from Eq.~\eqref{eq:dg} that
\begin{align}\label{eq:dg2}
&\int_0^T D_{x}g(t,X(t),H(t))^{\top}\sigma(X(t))d W(t)=g(T,X(T),H(T))-g(0,X(0),H(0))\nonumber\\
&\qquad\quad+\int_0^TL_{N+1}(H^{N+1}(t))\left\{\sum_{i=1}^{\bar{N}} b_i(1-K_i(H^{N+1}(t)))F_i(t,X(t)+w_{N+1},H^{N+1}(t))\right\}_+\nonumber\\
&\qquad\qquad\qquad\times X_{N+1}(t)(1-H_{N+1}(t))dt\nonumber\\
&\qquad\quad-\sum_{j=1}^{N+1}\int_0^T\left[g(t,{X}(t^-)+w_j,H^j(t^-))-g(t,{X}(t^-),H(t^-))\right]d M_j(t).
\end{align}
Then there exists a constant $C>0$ such that
\begin{align*}
&\Ex\left[\left|\int_0^T D_{x}g(t,X(t),H(t))^{\top}\sigma(X(t))\D W(t)\right|^2\right]\leq C \Ex\left[\left|g(T,X(T),H(T))-g(0,X(0),H(0))\right|^2\right]\nonumber\\
&\qquad+C\Ex\Bigg[\int_0^T\left(\sum_{i=1}^{\bar{N}} \left|b_i\right|\left[\left|F_i(t,X(t)+w_{N+1},H^{N+1}(t))\right|\right]\right)^2dt\left(\int_0^T X_{N+1}^2(t)dt\right)\Bigg]\nonumber\\
&\qquad+C\sum_{j=1}^{N+1}\Ex\left[\int_0^T\left|g(t,{X}(t)+w_j,H^j(t))-g(t,{X}(t),H(t))\right|^2X_j(t) dt \right].
\end{align*}
Notice that $(b_1,\ldots,b_{\bar{N}})$ is a finite sequence of real numbers. Using Proposition~\ref{lem:sol-Fi2} and thanks to Theorem~\ref{thm:sol}, it holds that there exists a constant $C>0$ such that
\begin{align*}
\Ex\left[\int_0^T\left|D_xg(t,X(t),H(t))^{\top} \sigma(X(t))\right|^2dt\right]&\leq C\left\{1 + \sum_{j=1}^{N+1}\Ex\left[\int_0^T X_{j}^2(t)dt\right]\right\}.
\end{align*}
Thus it follows from the estimate \eqref{eq:siest-2} that there exists a constant $C>0$ such that
\begin{align}\label{eq:siest-3}
\Ex\left[\int_0^{T\wedge\tau_{N+1}}\left|\theta^{GKW}(t)\right|^2d\left<Y_{N+1},Y_{N+1}\right>(t)\right]&\leq C\left\{1 + \sum_{j=1}^{N+1}\Ex\left[\int_0^TX_j^2(t)dt\right]\right\}.
\end{align}
Therefore, it suffices to estimate $\sum_{j=1}^{N+1}\Ex\left[\int_0^TX_j^2(t)dt\right]<+\infty$. Recall the default intensity process given by Eq.~\eqref{eq:hti}. Using It\^o's formula, it follows that for $j=1,\ldots,N+1$ and $t\in[0,T]$,
\begin{align*}
X_j^2(t) &= X_j^2(0) + 2\int_0^t X_j(s)\mu_j({ X(s)})ds +2\sum_{k=1}^K\int_0^t\sigma_{jk}({ X(s)})X_j(s)d W_k(s)\nonumber\\
&\quad+ \sum_{k=1}^K\int_0^t\sigma_{jk}^2({  X(s)})ds +\sum_{l=1}^{N+1}\int_0^t\left[\left(X_l(s^-)+w_{jl}\right)^2-X_l^2(s^-)\right]dH_l(s).
\end{align*}
Using the linear growth condition satisfied by $\mu(x)$ and $\sigma(x)$ in the assumption ({\bf A1}), it follows that there exists a constant $C>0$ such that for $j=1,\ldots,N+1$ and $t\in[0,T]$,
\begin{align*}
\Ex\left[X_j^2(t)\right]&\leq \Ex\left[X_j^2(0)\right]+C + C \int_0^t\Ex\left[X_j^2(s)\right]ds+{  C\sum_{j=1}^{N+1}\int_0^t\Ex\left[X_j^2(s)\right]ds }\nonumber\\
&\quad+\sum_{l=1}^{N+1}\int_0^t\Ex\left[\left(w_{jl}^2+2w_{jl}X_l(s)\right)X_l(s)\right]ds.
\end{align*}
For $j=1,\ldots,N+1$ and $t\in[0,T]$, this leads to the following inequalities
\begin{align*}
\sum_{j=1}^{N+1}\Ex\left[X_j^2(t)\right]&\leq \sum_{j=1}^{N+1}\Ex\left[X_j^2(0)\right] +C(N+1)+ { C(N+2)} \sum_{j=1}^{N+1}\int_0^t\Ex\left[X_j^2(s)\right]ds\nonumber\\
&\quad+C\sum_{l=1}^{N+1}\int_0^t\Ex\left[X_l^2(s)\right]ds + Ct.
\end{align*}
Gronwall's lemma implies that for all $t\in[0,T]$,
\begin{align*}
\sum_{j=1}^{N+1}\Ex\left[X_j^2(t)\right]&\leq \left\{C(T+N+1)+\sum_{j=1}^{N+1}\Ex\left[X_j^2(0)\right]\right\}e^{C(N+2)T}.
\end{align*}
Since the initial data $X_j(0)>0$ is square integrable for $j=1,\ldots,N+1$, this implies that
\begin{align}\label{eq:esti-Xjsum}
\sum_{j=1}^{N+1}\int_0^T\Ex\left[X_j^2(t)\right]dt&\leq \left\{C(T+N+1)+\sum_{j=1}^{N+1}\Ex\left[X_j^2(0)\right]\right\}Te^{C(N+2)T}<+\infty.
\end{align}
Hence, $\Ex\left[\int_0^{T\wedge\tau_{N+1}}\left|\theta^{GKW}(t)\right|^2d\left<Y_{N+1},Y_{N+1}\right>(t)\right]<+\infty$ using the estimate \eqref{eq:siest-3}. This completes the proof of the theorem. \hfill$\Box$

\section{Applications}\label{sec:applications}

We specialize the risk-minimizing strategy $\theta^{GKW} \in\Psi$ in the CDS contract referencing the counterparty ``$N+1$'' obtained in {Theorem}~\ref{thm:thetaGWK} to the case when the underlying traded portfolio consists of credit default swaps, risky bonds, or of a first-to-default claim. Recall the function $F_i(t,x,z)$ satisfying the recursive system of the backward Cauchy problems \eqref{eq:Cahchypro1i}, in which $\alpha=(1,1,1)$ for $i=1,\ldots,\bar{N}$ and $g(t,x,z)$ satisfies the recursive system \eqref{eq:F-K-g0}.

\subsection{Credit Swap Portfolio}

Recall the defaultable claim \eqref{eq:CDS}, i.e. $\bar{N}=N$ and for $i=1,\ldots,N+1$,
 \begin{align*}
  \xi_i=0,\quad a_i(t)=-\varepsilon_i,\quad Z_i(t)=L_i(t),\quad K_i(t)=H_i(t).
 \end{align*}
For $i=1,\ldots,N$, the recursive system \eqref{eq:Cahchypro1i} reduces to the Cauchy system~\eqref{eq:Cahchypro1CDS}, i.e., on $(t,x,z)\in[0,T)\times\R_+^{N+1}\times{\cal S}$,
\begin{align*}
&\left(\frac{\partial}{\partial t} + \A\right)F_{i}^{\rm cds}(t,x,z)-(1-z_i)\varepsilon_i-\sum_{j\neq i}z_i[L_i(z^j)-L_i(z)](1-z_j)x_j= 0
\end{align*}
with terminal condition $F_{i}^{\rm cds}(T,x,z)=L_i(z)z_i$.  On $(t,x,z)\in[0,T)\times\R_+^{N+1}\times{\cal S}$, the recursive Cauchy system \eqref{eq:F-K-g0}
\begin{align}\label{eq:F-K-g0CDS}
0&=\left(\frac{\partial}{\partial t}+{\cal A}\right)g^{\rm cds}(t,x,z)\\
&\quad+L_{N+1}(z^{N+1})\left\{\sum_{i=1}^{N} b_i(1-z_i)F_{i}^{\rm cds}(t,x+w_{N+1},z^{N+1})\right\}_+(1-z_{N+1})x_{N+1}\nonumber
\end{align}
with terminal condition $g^{\rm cds}(T,x,z)=0$. The unique risk-minimizing strategy is given by
\begin{align}\label{eq:thetaGKW200CDS}
\theta_{\rm cds}^{GKW}(t)&=\sum_{i=1}^3\frac{U_i^{\rm cds}(t,X(t^-),H(t^-))}{\Phi(t,X(t^-),H(t^-))},\qquad t\in\Ttau.
\end{align}
Above, for $(t,x,z)\in[0,T]\times\R_+^{N+1}\times{\cal S}$, the functions
\begin{align}\label{eq:coeffsCDS}
U_1^{\rm cds}(t,x,z)&:=\big<D_{x}g^{\rm cds}(t,x,z)^{\top}\sigma(x),V_{N+1}^{\rm cds}(t,x,z)^{\top}\sigma(x)\big>;\nonumber\\
U_2^{\rm cds}(t,x,z)&:=L_{N+1}(z^{N+1})\left\{\sum_{i=1}^{N} b_i(1-z_i)F_i^{\rm cds}(t,x+w_{N+1},z^{N+1})\right\}_+ G_{N+1,N+1}^{\rm cds}(t,x,z)x_{N+1};\nonumber\\
U_3^{\rm cds}(t,x,z)&:=\sum_{j=1}^{N+1}\left[g^{\rm cds}(t,x+w_j,z^j)-g^{\rm cds}(t,x,z)\right]\\
&\quad\times\big\{G_{N+1,j}^{\rm cds}(t,x,z)-z_{N+1}[L_{N+1}(z^j)-L_{N+1}(z)]\big\}x_j(1-z_j).\nonumber
\end{align}

Consider a portfolio consisting of a single name CDS, that is $N=1$, traded against the risky counterparty ``$2$'' of the investor. In this case, we obtain closed-form solutions for the two types of recursive Cauchy systems. Using these closed-form solutions, one can derive the risk-minimizing strategy $\theta_{\rm cds}^{GKW}(t)$ using \eqref{eq:thetaGKW200CDS}. We distinguish the following cases:
\begin{itemize}
  \item $z=(1,1)$. We have $F_i^{\rm cds}(t,x,(1,1))=L_i((1,1))$ for $i=1,2$ and $g^{\rm cds}(t,x,(1,1))=0$.
  \item $z=(1,0)$. We have $g^{\rm cds}(t,x,(1,0))=0$ and
  \begin{align*}
  F_1^{\rm cds}(t,x,(1,0))&=L_1((1,0))\Ex\left[e^{-\int_t^T\tilde{X}_2^{(t,x)}(s)ds}\right]
+L_1((1,0))\Ex\left[\int_t^T\tilde{X}_2^{(t,x)}(s)e^{-\int_t^s\tilde{X}_2^{(t,x)}(u)du}ds\right];\nonumber\\
  F_2^{\rm cds}(t,x,(1,0))&=\Ex\left[\int_t^T\big\{L_2((1,1))\tilde{X}_2^{(t,x)}(s)-\varepsilon_2\big\}e^{-\int_t^s\tilde{X}_2^{(t,x)}(u)du}ds\right].
  \end{align*}
  \item $z=(0,1)$. We have $g^{\rm cds}(t,x,(0,1))=0$ and
  \begin{align*}
  F_1^{\rm cds}(t,x,(0,1))&=\Ex\left[\int_t^T\big\{L_1((1,1))\tilde{X}_1^{(t,x)}(s)-\varepsilon_1\big\}e^{-\int_t^s\tilde{X}_1^{(t,x)}(u)du}ds\right];\nonumber\\
F_2^{\rm cds}(t,x,(0,1))&=L_2((0,1))\Ex\left[e^{-\int_t^T\tilde{X}_1^{(t,x)}(s)ds}\right]
+L_2((0,1))\Ex\left[\int_t^T\tilde{X}_1^{(t,x)}(s)e^{-\int_t^s\tilde{X}_1^{(t,x)}(u)du}ds\right].
  \end{align*}
  \item $z=(0,0)$. We have
 \begin{align*}
  g^{\rm cds}(t,x,(0,0))&=L_2((0,1))\Ex\Bigg[\int_t^T\tilde{X}_2^{(t,x)}(s)\left\{b_1F_1^{\rm cds}(s,\tilde{X}^{(t,x)}(s)+w_2,(0,1))\right\}_+\nonumber\\
&\qquad\times e^{-\int_t^s(\tilde{X}_1^{(t,x)}(u)+\tilde{X}_2^{(t,x)}(u))du}ds\Bigg],
 \end{align*}
and
\begin{align*}
F_i^{\rm cds}(t,x,(0,0))&=\Ex\left[\int_t^T\left(\sum_{j=1}^2F_{i}^{\rm cds}(s,\tilde{X}^{(t,x)}(s),(0,0)^j)\tilde{X}_j^{(t,x)}(s)-\varepsilon_i\right)e^{-\int_t^s(\tilde{X}_1^{(t,x)}(u)+\tilde{X}_2^{(t,x)}(u))du}ds\right].
\end{align*}
\end{itemize}

\subsection{Risky Bonds Portfolio}

Recall the default claim \eqref{eq:bond} associated with the risky bonds portfolio, i.e. $\bar{N}=N$ and for $i=1,\ldots,N$,
\begin{align*}
  \xi_i=1,\quad a_i(t)=\varepsilon_i,\quad Z_i(t)=1-L_i(t),\quad K_i(t)=H_i(t),
\end{align*}
{while for the counterpary
\begin{align*}
  \xi_{N+1}=0,\quad a_{N+1}(t)=-\varepsilon_{N+1},\quad Z_{N+1}(t)=L_{N+1}(t),\quad K_{N+1}(t)=H_{N+1}(t).
\end{align*} }
Then for $i=1,\ldots,N$, the recursive system \eqref{eq:Cahchypro1i} reduces to the following Cauchy system given by, on $(t,x,z)\in[0,T)\times\R_+^{N+1}\times{\cal S}$,
\begin{align}\label{eq:Cahchypro1irisky-bond}
&\left(\frac{\partial}{\partial t} + \A\right)F_{i}^{\rm bond}(t,x,z)+(1-z_i)\varepsilon_i+\sum_{j\neq i}z_i[L_i(z^j)-L_i(z)](1-z_j)x_j= 0
\end{align}
with terminal condition $F_{i}^{\rm bond}(T,x,z)=(1-z_i)+(1-L_i(z))z_i$. The recursive Cauchy system \eqref{eq:F-K-g0} is reduced to, on $(t,x,z)\in[0,T)\times\R_+^{N+1}\times{\cal S}$,
\begin{align}\label{eq:F-K-g0bond}
0&=\left(\frac{\partial}{\partial t}+{\cal A}\right)g^{\rm bond}(t,x,z)\\
&\quad+L_{N+1}(z^{N+1})\left\{\sum_{i=1}^{N} b_i(1-z_i)F_{i}^{\rm bond}(t,x+w_{N+1},z^{N+1})\right\}_+(1-z_{N+1})x_{N+1}\nonumber
\end{align}
with terminal condition $g^{\rm bond}(T,x,z)=0$. The unique risk-minimizing strategy on risky bonds portfolio is given by
\begin{align}\label{eq:thetaGKW200bond}
\theta_{\rm bond}^{GKW}(t)&=\sum_{i=1}^3\frac{U_i^{\rm bond}(t,X(t^-),H(t^-))}{\Phi(t,X(t^-),H(t^-))},\qquad t\in\Ttau.
\end{align}
Above, for $(t,x,z)\in[0,T]\times\R_+^{N+1}\times{\cal S}$, the functions
\begin{align}\label{eq:coeffsbond}
U_1^{\rm bond}(t,x,z)&:=\big<D_{x}g^{\rm bond}(t,x,z)^{\top}\sigma(x),V_{N+1}^{\rm cds}(t,x,z)^{\top}\sigma(x)\big>;\nonumber\\
U_2^{\rm bond}(t,x,z)&:=L_{N+1}(z^{N+1})\left\{\sum_{i=1}^{N} b_i(1-z_i)F_i^{\rm bond}(t,x+w_{N+1},z^{N+1})\right\}_+ G_{N+1,N+1}^{\rm cds}(t,x,z)x_{N+1};\nonumber\\
U_3^{\rm bond}(t,x,z)&:=\sum_{j=1}^{N+1}\left[g^{\rm bond}(t,x+w_j,z^j)-g^{\rm bond}(t,x,z)\right]\\
&\quad\times\big\{G_{N+1,j}^{\rm cds}(t,x,z)-z_{N+1}[L_{N+1}(z^j)-L_{N+1}(z)]\big\}x_j(1-z_j).\nonumber
\end{align}

Consider a portfolio consisting of a single name risky bond, that is $N=1$, traded against the risky counterparty ``$2$'' of the investor. Again, the two types of recursive Cauchy systems admits closed-form solutions, and thus allows us to derive the risk-minimizing strategy $\theta_{\rm cds}^{GKW}(t)$ using \eqref{eq:thetaGKW200bond}. We consider the following cases:
\begin{itemize}
  \item $z=(1,1)$. We have $F_i^{\rm bond}(t,x,(1,1))=1-L_i((1,1))$ for $i=1,2$ and $g^{\rm bond}(t,x,(1,1))=0$.
  \item $z=(1,0)$. We have $g^{\rm bond}(t,x,(1,0))=0$ and
  \begin{align*}
  F_1^{\rm bond}(t,x,(1,0))&=(1-L_1((1,0)))\Ex\left[e^{-\int_t^T\tilde{X}_2^{(t,x)}(s)ds}\right]\nonumber\\
&\quad+(1-L_1((1,0)))\Ex\left[\int_t^T\tilde{X}_2^{(t,x)}(s)e^{-\int_t^s\tilde{X}_2^{(t,x)}(u)du}ds\right].
  \end{align*}
  \item $z=(0,1)$. We have $g^{\rm bond}(t,x,(0,1))=0$ and
  \begin{align*}
  F_1^{\rm bond}(t,x,(0,1))&=\Ex\left[e^{-\int_t^T\tilde{X}_1^{(t,x)}(u)du}\right]+\Ex\left[\int_t^T\big\{(1-L_1((1,1)))\tilde{X}_1^{(t,x)}(s)
+\varepsilon_1\big\}e^{-\int_t^s\tilde{X}_1^{(t,x)}(u)du}ds\right].
  \end{align*}
  \item $z=(0,0)$. We have
 \begin{align*}
  g^{\rm bond}(t,x,(0,0))&=L_2((0,1))\Ex\Bigg[\int_t^T\tilde{X}_2^{(t,x)}(s)\left\{b_1F_1^{\rm bond}(s,\tilde{X}^{(t,x)}(s)+w_2,(0,1))\right\}_+\nonumber\\
&\qquad\times e^{-\int_t^s(\tilde{X}_1^{(t,x)}(u)+\tilde{X}_2^{(t,x)}(u))du}ds\Bigg],
 \end{align*}
and
\begin{align*}
&F_1^{\rm bond}(t,x,(0,0))=\Ex\left[e^{-\int_t^T(\tilde{X}_1^{(t,x)}(u)+\tilde{X}_2^{(t,x)}(u))du}ds\right]\nonumber\\
&\qquad\quad+\Ex\left[\int_t^T\left(\sum_{j=1}^2F_{1}^{\rm bond}(s,\tilde{X}^{(t,x)}(s),(0,0)^j)\tilde{X}_j^{(t,x)}(s)+\varepsilon_1\right)e^{-\int_t^s(\tilde{X}_1^{(t,x)}(u)+\tilde{X}_2^{(t,x)}(u))du}ds\right].
\end{align*}
\end{itemize}

\subsection{First-to-Default Claim}

Recall the first-to-default claim given in~\eqref{eq:first-to}, i.e. $\bar{N}=1$ and for $i=1,2$,
\begin{align*}
  &\xi_1=0,\quad a_1(t)=\varepsilon,\quad Z_1(t)=\sum_{i=1}^N L_i(t)H_i(t),\quad K_1(t)=1-\prod_{i=1}^N(1-H_i(t));\\
   &\xi_2=0,\quad a_2(t)=-\varepsilon_{N+1},\quad Z_2(t)=L_{N+1}(t),\quad K_2(t)=H_{N+1}(t).\nonumber
 \end{align*}
The recursive system \eqref{eq:Cahchypro1i} reduces to the following Cauchy system given by, on $(t,x,z)\in[0,T)\times\R_+^{N+1}\times{\cal S}$,
\begin{align}\label{eq:Cahchypro1iftd}
&\left(\frac{\partial}{\partial t} + \A\right)F_{1}^{\rm ftd}(t,x,z)+\varepsilon\prod_{i=1}^N(1-z_i)-\sum_{j=1}^{N+1}K_1(z)[Z_1(z^j)-Z_1(z)](1-z_j)x_j= 0
\end{align}
where
\begin{align}\label{eq:3rdtermftdappl}
&\sum_{j=1}^{N+1}K_1(z)[Z_1(z^j)-Z_1(z)](1-z_j)x_j\nonumber\\
&\quad=\left(1-\prod_{i=1}^N(1-z_i)\right)\sum_{j=1}^{N+1}(1-z_j)x_j\sum_{i=1}^N\big[L_i(z^j)z_i{\bf1}_{j\neq i}+L_i(z^i)(1-z_i)-L_i(z)z_i\big].
\end{align}
The terminal condition is given by
\begin{align}\label{eq:terminal-conditioniftdappl}
F_{1}^{{\rm ftd}}(T,x,z)=\left(\sum_{i=1}^N L_i(z)z_i\right) \left(1-\prod_{i=1}^N(1-z_i)\right).
\end{align}
The recursive Cauchy system \eqref{eq:F-K-g0} is reduced to, on $(t,x,z)\in[0,T)\times\R_+^{N+1}\times{\cal S}$,
\begin{align}\label{eq:F-K-g0ftd}
0&=\left(\frac{\partial}{\partial t}+{\cal A}\right)g^{\rm ftd}(t,x,z)\nonumber\\
&\quad+L_{N+1}(z^{N+1})\left\{b_1(1-K_1(z^{N+1}))F_{1}^{\rm ftd}(t,x+w_{N+1},z^{N+1})\right\}_+(1-z_{N+1})x_{N+1}
\end{align}
with terminal condition $g^{\rm ftd}(T,x,z)=0$. The unique risk-minimizing strategy of the CVA on the first-to-default claim is given by
\begin{align}\label{eq:thetaGKW200ftd}
\theta_{\rm ftd}^{GKW}(t)&=\sum_{i=1}^3\frac{U_i^{\rm ftd}(t,X(t^-),H(t^-))}{\Phi(t,X(t^-),H(t^-))},\qquad t\in\Ttau.
\end{align}
Above, for $(t,x,z)\in[0,T]\times\R_+^{N+1}\times{\cal S}$, the functions
\begin{align}\label{eq:coeffsftd}
U_1^{\rm ftd}(t,x,z)&:=\big<D_{x}g^{\rm ftd}(t,x,z)^{\top}\sigma(x),V_{N+1}^{\rm cds}(t,x,z)^{\top}\sigma(x)\big>;\nonumber\\
U_2^{\rm ftd}(t,x,z)&:=L_{N+1}(z^{N+1})\big\{b_1(1-K_1(z))F_1^{\rm ftd}(t,x+w_{N+1},z^{N+1})\big\}_+ G_{N+1,N+1}^{\rm cds}(t,x,z)x_{N+1};\nonumber\\
U_3^{\rm ftd}(t,x,z)&:=\sum_{j=1}^{N+1}\left[g^{\rm ftd}(t,x+w_j,z^j)-g^{\rm ftd}(t,x,z)\right]\\
&\quad\times\big\{G_{N+1,j}^{\rm cds}(t,x,z)-z_{N+1}[L_{N+1}(z^j)-L_{N+1}(z)]\big\}x_j(1-z_j).\nonumber
\end{align}

Consider a first-to-default claim in a basket of two names, that is $N=2$, traded against the risky counterparty ``$3$'' of the investor. Both types of recursive Cauchy systems can be solved in closed-form, and the risk-minimizing strategy $\theta_{\rm cds}^{GKW}(t)$ can then be computed using Eq.~\eqref{eq:thetaGKW200ftd}.
We have $\bar{\tau}_1=\tau_1\wedge\tau_2$ and $\bar{\tau}_2=\tau_{3}$. We separately treat the following cases:
\begin{itemize}
  \item $z=(1,1,1)$. We have $F_1^{\rm ftd}(t,x,(1,1,1))=L_1((1,1,1))+L_2((1,1,1))$ and $g^{\rm ftd}(t,x,(1,1,1))=0$.
\item $z=(1,1,0)$. We have
\begin{align*}
g^{\rm ftd}(t,x,(1,1,0))&=L_3((1,1,1))\left(\sum_{i=1}^2L_i((1,1,1))\right)\{b_1\}_+\Ex\left[\int_t^T\tilde{X}_3^{(t,x)}(s)e^{-\int_t^s\tilde{X}_3^{(t,x)}(u)du}\right],
\end{align*}
and
\begin{align*}
F_1^{\rm ftd}(t,x,(1,1,0))&=\left(\sum_{i=1}^2L_i((1,1,0))\right)\left\{\Ex\left[e^{-\int_t^T\tilde{X}_3^{(t,x)}(s)ds}\right]+
\Ex\left[\int_t^T\tilde{X}_3^{(t,x)}(s)e^{-\int_t^s\tilde{X}_3^{(t,x)}(u)du}\right]\right\}.
\end{align*}
\item $z=(1,0,1)$. We have $g^{\rm ftd}(t,x,(1,0,1))=0$ and
\begin{align*}
F_1^{\rm ftd}(t,x,(1,0,1))&=L_1((1,0,1))\left\{\Ex\left[e^{-\int_t^T\tilde{X}_2^{(t,x)}(s)ds}\right]
+\Ex\left[\int_t^T\tilde{X}_2^{(t,x)}(s)e^{-\int_t^s\tilde{X}_2^{(t,x)}(u)du}\right]\right\}.
\end{align*}
\item $z=(0,1,1)$. We have $g^{\rm ftd}(t,x,(0,1,1))=0$ and
\begin{align*}
F_1^{\rm ftd}(t,x,(0,1,1))&=L_2((0,1,1))\left\{\Ex\left[e^{-\int_t^T\tilde{X}_1^{(t,x)}(s)ds}\right]
+\Ex\left[\int_t^T\tilde{X}_1^{(t,x)}(s)e^{-\int_t^s\tilde{X}_1^{(t,x)}(u)du}\right]\right\}.
\end{align*}
\item $z=(1,0,0)$. We have
\begin{align*}
g^{\rm ftd}(t,x,(1,0,0))&=\Ex\left[\int_t^T\tilde{X}_2^{(t,x)}(s)g^{\rm ftd}(s,\tilde{X}^{(t,x)}(s),(1,1,0))e^{-\int_t^s(\tilde{X}_1^{(t,x)}(u)+\tilde{X}_2^{(t,x)}(u))du}\right],
\end{align*}
and
\begin{align*}
&F_1^{\rm ftd}(t,x,(1,0,0))=L_1((1,0,0))\Ex\left[e^{-\int_t^T(\tilde{X}_2^{(t,x)}(s)+\tilde{X}_3^{(t,x)}(s))ds}\right]\nonumber\\
&\quad+\Ex\Bigg[\int_t^T\Big\{\tilde{X}_2^{(t,x)}(s)\big(F_1^{\rm ftd}(s,\tilde{X}^{(t,x)}(s),(1,1,0))-L_1((1,1,0))-L_2((1,1,0))+L_1((1,0,0))\big)\nonumber\\
&\qquad+\tilde{X}_3^{(t,x)}(s)\big(F_1^{\rm ftd}(s,\tilde{X}^{(t,x)}(s),(1,0,1))-L_1((1,0,1))-L_2((1,1,0))+L_1((1,0,0))\big)\Big\}\nonumber\\
&\qquad\qquad\times e^{-\int_t^s(\tilde{X}_2^{(t,x)}(u)+\tilde{X}_3^{(t,x)}(u))du}\Bigg].
\end{align*}
\item $z=(0,1,0)$. We have
\begin{align*}
g^{\rm ftd}(t,x,(0,1,0))&=\Ex\left[\int_t^T\tilde{X}_1^{(t,x)}(s)g^{\rm ftd}(s,\tilde{X}^{(t,x)}(s),(1,1,0))e^{-\int_t^s(\tilde{X}_1^{(t,x)}(u)+\tilde{X}_3^{(t,x)}(u))du}\right],
\end{align*}
and
\begin{align*}
&F_1^{\rm ftd}(t,x,(0,1,0))=L_2((0,1,0))\Ex\left[e^{-\int_t^T(\tilde{X}_1^{(t,x)}(s)+\tilde{X}_3^{(t,x)}(s))ds}\right]\nonumber\\
&\quad+\Ex\Bigg[\int_t^T\Big\{\tilde{X}_1^{(t,x)}(s)\big(F_1^{\rm ftd}(s,\tilde{X}^{(t,x)}(s),(1,1,0))-L_1((1,1,0))-L_2((1,1,0))+L_2((0,1,0))\big)\nonumber\\
&\qquad+\tilde{X}_3^{(t,x)}(s)\big(F_1^{\rm ftd}(s,\tilde{X}^{(t,x)}(s),(0,1,1))-L_1((1,1,0))-L_2((0,1,1))+L_2((0,1,0))\big)\Big\}\nonumber\\
&\qquad\qquad\times e^{-\int_t^s(\tilde{X}_1^{(t,x)}(u)+\tilde{X}_3^{(t,x)}(u))du}\Bigg].
\end{align*}
\item $z=(0,0,1)$. We have $g^{\rm ftd}(t,x,(0,0,1))=0$ and
\begin{align*}
&F_1^{\rm ftd}(t,x,(0,0,1))=\Ex\Bigg[\int_t^T\Big\{\tilde{X}_1^{(t,x)}(s)F_1^{\rm ftd}(s,\tilde{X}^{(t,x)}(s),(1,0,1))\nonumber\\
&\qquad+\tilde{X}_2^{(t,x)}(s)F_1^{\rm ftd}(s,\tilde{X}^{(t,x)}(s),(0,1,1))\Big\}e^{-\int_t^s(\tilde{X}_1^{(t,x)}(u)+\tilde{X}_2^{(t,x)}(u))du}\Bigg].
\end{align*}
\item $z=(0,0,0)$. We have
\begin{align*}
&g^{\rm ftd}(t,x,(0,0,0))=\Ex\Bigg[\int_t^T\Big\{\tilde{X}_1^{(t,x)}(s)g_1^{\rm ftd}(s,\tilde{X}^{(t,x)}(s),(1,0,0))+\tilde{X}_2^{(t,x)}(s)g^{\rm ftd}(s,\tilde{X}^{(t,x)}(s),(0,1,0))\nonumber\\
&\quad+L_3((0,0,1))\tilde{X}_3^{(t,x)}(s)\big\{b_1F_1^{\rm ftd}(t,x+w_3,(0,0,1))\big\}_+\Big\} e^{-\int_t^s(\tilde{X}_1^{(t,x)}(u)+\tilde{X}_2^{(t,x)}(u)+\tilde{X}_3^{(t,x)}(u))du}\Bigg],
\end{align*}
and
\begin{align*}
&F_1^{\rm ftd}(t,x,(0,0,0))=\Ex\Bigg[\int_t^T\Big\{\tilde{X}_1^{(t,x)}(s)F_1^{\rm ftd}(s,\tilde{X}^{(t,x)}(s),(1,0,0))+\tilde{X}_2^{(t,x)}(s)F_1^{\rm ftd}(s,\tilde{X}^{(t,x)}(s),(0,1,0))\nonumber\\
&\qquad+\tilde{X}_3^{(t,x)}(s)F_1^{\rm ftd}(s,\tilde{X}^{(t,x)}(s),(0,0,1))+\varepsilon\Big\}e^{-\int_t^s(\tilde{X}_1^{(t,x)}(u)+\tilde{X}_2^{(t,x)}(u)+\tilde{X}_3^{(t,x)}(u))du}\Bigg].
\end{align*}
\end{itemize}

{The probabilistic representation of the above quantities makes it possible to develop efficient Monte-Carlo simulation methods to approximate the risk-minimizing hedging strategy. }

\section{Conclusion}\label{sec:conclusion}
In this paper, we have studied dynamic hedging of counterparty risk of defaultable claims. We have considered a market model consisting of securities whose payments are contingent to the default events of $N$ firms. We have carried out our analysis under a model of direct default contagion, in which the intensities follow jump diffusion processes, and interact with each other through their common dependence on the default state of the portfolio. Consistently with market practice, we have used the liquidly traded CDS contract written on the risky investor's counterparty to hedge against the credit valuation adjustment claim. We have derived hedging strategies in the risk-minimizing sense, i.e. using a hedging method which keep the replicability constraint of the claim and at the same time guarantees that the self-financing condition is satisfied on average. We have shown that the hedging strategy is given by the integrand of the GKW decomposition for the CVA payment stream (see Theorem~\ref{thm:thetaGWK}), and admits an explicit representation in terms of solutions to a non-linear system of backward Cauchy problems. We  have established the existence of a unique {smooth} solution to this system, defined on an unbounded domain and having non-Lipschitz coefficients, by proving the uniform integrability of the family generated by the corresponding Feymann-Kac's representations. Due to its analytical tractability, our framework can be used to support decisions of risk management desks within financial firms, dealing with the critical problem of counterparty risk hedging.

\appendix

\section{Proofs} \label{sec:techproofs}


\noindent{\bf Proof of Lemma~\ref{lem:Dfirst-todefault}.}\quad By virtue of Definition~\ref{def:dividend}, we have the representation of the dividend process of the first-to-default claim  given by
\begin{align}\label{eq:dividend-ftd}
D(t) = -\varepsilon\int_{0}^{t \wedge T} (1-K(u))du + \sum_{i=1}^N\int_{0}^{t \wedge T} L_i(H(u))H_i(u) d K(u).
\end{align}
The third term of the above dividend process is in fact given by
\begin{align*}
\sum_{i=1}^N\int_{0}^{t \wedge T} L_i(H(u))H_i(u) d K(u) =\sum_{i=1}^NL_i(H(\bar{\tau}_1))H_i(\bar{\tau}_1){\bf1}_{\bar{\tau}_1\leq t\wedge T}=\sum_{i=1}^NL_i(H(\bar{\tau}_1)){\bf1}_{\tau_i\leq \bar{\tau}_1}{\bf1}_{\bar{\tau}_1\leq t\wedge T}.
\end{align*}
Notice that for all $i=1,\ldots,N$, we have $\tau_i\geq \bar{\tau}_1=\tau_1\wedge\cdots\wedge\tau_N$, a.s.. Hence ${\bf1}_{\tau_i\leq \bar{\tau}_1}={\bf1}_{\tau_i=\bar{\tau}_1}$, a.s.. Thus the above equality becomes that
\begin{align*}
\sum_{i=1}^N\int_{0}^{t \wedge T} L_i(H(u))H_i(u) d K(u) &=\sum_{i=1}^NL_i(H(\bar{\tau}_1)){\bf1}_{\tau_i\leq \bar{\tau}_1}{\bf1}_{\bar{\tau}_1\leq t\wedge T}=\sum_{i=1}^NL_i(H(\bar{\tau}_1)){\bf1}_{\tau_i=\bar{\tau}_1}{\bf1}_{\bar{\tau}_1\leq t\wedge T}.
\end{align*}
This results in the dividend representation given by Eq.~\eqref{eq:Dfirst}. \hfill$\Box$\\

\noindent{\bf Proof of Proposition~\ref{lem:sol-Fi2}.} \quad On $(t,x)\in[0,T)\times\R_+^{N+1}$, we rewrite the problem~\eqref{eq:Cahchyprol} in the following abstract linear form:
\begin{align}\label{eq:PDEFi2000}
&\left(\frac{\partial}{\partial t}+\tilde{\cal A}\right)u(t,x)+h(x)u(t,x)+w(t,x)=0
\end{align}
with terminal condition $u(T,x)=\alpha_1\xi^{(l)}(1-K^{(l)})+\alpha_2Z^{(l)}K^{(l)}$ for all $x\in\R_+^{N+1}$. On $(t,x)\in\R_+^{N+1}$, the coefficients
\begin{align*}
h(x) &:= -\sum_{j\notin\{j_1,\ldots,j_l\}}x_j,\nonumber\\
w(t,x)&:=\sum_{j\notin\{j_1,\ldots,j_l\}}x_j\big[F_{\alpha}^{(l+1),j}(t,x+w_j)-\alpha_3K^{(l)}(Z^{(l+1),j}-Z^{(l)})\big]+\alpha_3(1-K^{(l)})a^{(l)}.
\end{align*}
We will apply Theorem 1 of \cite{health2000} to prove existence and uniqueness of classical solutions to Eq.~\eqref{eq:PDEFi2000} by verifying that their imposed conditions [A1], [A2], [A3'] and [A3a']-[A3e'] hold in our case. Consider a sequence of bounded domains $D_n:=(\frac{1}{n},n)^{N+1}$, $n\in\N$, with smoothed corners and satisfying $\cup_{n=1}^{\infty}D_n=\R_+^{N+1}$. Thus we verify that the condition [A3'] on the domain of the equation holds. By the assumptions ({\bf A1}) and ({\bf A2}), the conditions [A1] and [A2] for the coefficients $\mu(x)$ and $\sigma(x)$ can be satisfied. This also implies that [A3a'] holds. Moreover, since $\sigma\sigma^{\top}(x)$ is continuous and invertible under assumptions ({\bf A1}) and ({\bf A2}), $\sigma\sigma^{\top}(x)$ is uniformly elliptic on $(t,x)\times\overline{D}_n$, i.e. [A3b'] holds.
Notice that $F_{\alpha}^{(l+1),j}(t,x+w_j)$ is bounded and $C^{1,2}$ in $(t,x)$ by the induction hypothesis. Additionally, notice that $h(x)$ is linear in $x$. Then the conditions [A3c'] and [A3d'] on the coefficients $h(x)$ and $w(t,x)$ on $(t,x)\in[0,T]\times\overline{D}_n$ are satisfied. Finally we need to verify [A3e']. For this, it suffices to prove the uniform integrability of the family
\begin{align}\label{eq:ui}
\left\{\int_t^Tw(s,\tilde{X}^{(t,x)}(s))e^{-\int_t^sh(\tilde{X}^{(t,x)}(u))du}ds;\ (t,x)\in[0,T]\times\R_+^{N+1}\right\}.
\end{align}
In fact, by the inductive hypothesis that $F_{\alpha}^{(l+1),j}(t,x)$ is nonnegative and bounded on $[0,T]\times\R_+^{N+1}$ for all $j\notin\{j_1,\ldots,j_l\}$, there exists a constant $C>0$ independent of $(t,x)$ such that for all $(t,x)\in[0,T]\times\R_+^{N+1}$ we have
\begin{align}\label{eq:ui2}
&\Ex\left[\left|\int_t^Tw(s,\tilde{X}^{(t,x)}(s))e^{\int_t^sh(\tilde{X}^{(t,x)}(u))du}ds\right|^2\right]\nonumber\\
&\quad\leq C\Ex\left[\left|\int_t^Te^{-\int_t^s(\sum_{k\notin\{j_1,\ldots,j_l\}} \tilde{X}_k^{(t,x)}(u))du}\left(1+\sum_{j\notin\{j_1,\ldots,j_l\}}\tilde{X}_j^{(t,x)}(s)\right)ds\right|^2 \right]\nonumber\\
&\quad\leq 2CT^2+2C\Ex\left[\left|\int_t^Te^{-\int_t^s(\sum_{k\notin\{j_1,\ldots,j_l\}} \tilde{X}_k^{(t,x)}(u))du}d\left(\int_t^s\sum_{j\notin\{j_1,\ldots,j_l\}}\tilde{X}_j^{(t,x)}(u)du\right)\right|^2\right]\nonumber\\
&\quad\leq 2CT^2+2C\left\{1+\left|\Ex\left[e^{-\int_t^T(\sum_{k\notin\{j_1,\ldots,j_l\}} \tilde{X}_k^{(t,x)}(u))du}\right]\right|^2\right\}\nonumber\\
&\quad\leq 2CT^2+4C.
\end{align}
This yields the existstence of a constant $C>0$, independent of $(t,x)$, such that
\begin{align*}
\sup_{(t,x)\in[0,T]\times\R_+^{N+1}}\Ex\left[\left|\int_t^Tw(s,\tilde{X}^{(t,x)}(s))e^{\int_t^sh(\tilde{X}^{(t,x)}(u))du}ds\right|^2\right]\leq C<+\infty.
\end{align*}
This yields the uniform integrability of the family \eqref{eq:ui}. It implies the condition [A3e'] of \cite{health2000} is satisfied. Using Theorem 1 of \cite{health2000}, Eq.~\eqref{eq:PDEFi2000} admits a unique classical solution $u(t,x)$ on $[0,T]\times\R_+^{N+1}$. Moreover this solution admits the probabilistic representation given by \eqref{eq:FY-Fi2}. Further the estimate~\eqref{eq:ui2} implies that the probabilistic representation given by \eqref{eq:FY-Fi2} is bounded for all $(t,x)\in[0,T]\times\R_+^{N+1}$.  This completes the proof of the proposition. \hfill$\Box$\\

\noindent{\bf Proof of Lemma~\ref{lem:gain}.} \quad It follows from Eq.~\eqref{eq:dividendT} that
\begin{align*}
D(T) = \xi(H(T))(1-K(T)) +\int_{0}^{T} (1-K(u)) a(u) du + \int_{0}^{T} Z(u) d K(u).
\end{align*}
Using integration by parts \eqref{eq:inteparts}, we have that
\begin{align*}
D(T) &= \xi(H(T))(1-K(H(T))) + \int_{0}^{T} (1-K(H(u))) a(u) du + Z(H(T))K(H(T))\nonumber\\
&\quad-Z(H(0))K(H(0))-\int_{0}^{T} K(H(u^-)) d Z(H(u)).
\end{align*}
Since $K(0)=0$, it follows from Proposition~\ref{lem:sol-Fi2} that
\begin{align}\label{eq:Ysum}
Y(t) &= F_{(1,1,1)}(t,X(t),H(t))+\int_{0}^{t} (1-K(u)) a(u) du -\int_{0}^{t} K(H(u^-)) d Z(H(u)).
\end{align}
Above, $F_{(1,1,1)}(t,x,z)$ is the unique bounded classical solution to the recursive system of the backward Cauchy problems given by, on $(t,x,z)\in[0,T)\times\R_+^{N+1}\times{\cal S}$,
\begin{align}\label{eq:Cahchypro1111}
&\left(\frac{\partial}{\partial t} + \A\right)F_{(1,1,1)}(t,x,z)+(1-K(z))a(z)-\sum_{j=1}^{N+1}K(z)[Z(z^j)-Z(z)](1-z_j)x_j= 0
\end{align}
with terminal condition
\begin{align}\label{eq:terminal-condition111}
F_{(1,1,1)}(T,x,z)=\xi(z)(1-K(z))+Z(z)K(z),\qquad (x,z)\in \R_+^{N+1}\times{\cal S}.
\end{align}
Applying It\^o's formula, we obtain
\begin{align*}
&F_{(1,1,1)}(t,X(t),H(t)) =F_{(1,1,1)}(0,X(0),H(0))\nonumber\\
&\qquad +\int_0^t\left\{\sum_{j=1}^{N+1}K(H(u))[Z(H^j(u))-Z(H(u))](1-H_j(u))X_j(u)-(1-K(H(u)))a(H(u))\right\}du\nonumber\\
&\qquad+ \int_0^t D_xF_{(1,1,1)}(u,X(u),H(u))^{\top}\sigma(X(u))dW(u)\nonumber\\
&\qquad+\sum_{j=1}^{N+1} \int_0^t [F_{(1,1,1)}(u,X(u^-)+w_j,H^j(u^-))-F_{(1,1,1)}(u,X(u^-),H(u^-))]dM_j(u).
\end{align*}
Using Eq.~\eqref{eq:Ysum}, we deduce
\begin{align*}
dY(t)&=D_xF_{(1,1,1)}(t,X(t),H(t))^{\top}\sigma(X(t))dW(t)\nonumber\\
&\quad+\sum_{j=1}^{N+1}[F_{(1,1,1)}(t,X(t^-)+w_j,H^j(t^-))-F_{(1,1,1)}(t,X(t^-),H(t^-))]dM_j(t)\nonumber\\
&\quad-\sum_{j=1}^{N+1}K(H(t^-))[Z(H^j(t^-))-Z(H(t^-))]dM_j(t).
\end{align*}
This yields the dynamics \eqref{eq:Yi} of the gain process. \hfill$\Box$ \\

\noindent{\bf Proof of Proposition~\ref{thm:strategy}.}\quad
{The proof is an easy extension of that of Theorem 2.4 in \cite{guided} to the case of payment stream with random delivery state.
Let $\varphi^* = (\theta^*, \eta^*)$ be the strategy defined in Eq.~\eqref{eq:thetaGKW}. Then $V^{\varphi^*}(t) :=  \theta^*(t) Y_{N+1}(t)  + \eta^*(t) = V(t) - \Theta(t)$
and hence from Eq.s~(\ref{eq:cost}) and (\ref{eq:GKWbis}), it holds that
\begin{align*}
C^{\varphi^*}(t) =  V(t)  -   \int_0^t \theta^{GKW}(u) dY_{N+1}(u) = \Ex \left[  \Theta(T\wedge\tau_{N+1}) \right] + A(t)
\end{align*}
which implies that $C^{\varphi^*}$ turns out to be a $\Gx$-martingale strongly orthogonal to $Y_{N+1}$. By virtue of Lemma 2.3 of \cite{guided}, we can restrict ourself to consider mean self-financing strategies. Let $\varphi = (\theta, \eta)\in \Psi$ be a mean self-financing strategy $0$-achieving, i.e. $V^{\varphi}(\tau_{N+1} \wedge T)=0$, then from Eq.~\eqref{eq:cost}, it follows that
\begin{align}\label{conti}
C^{\varphi}(T\wedge\tau_{N+1}) - C^{\varphi}(t) = \Theta(T\wedge\tau_{N+1}) -\Theta(t) - V^\varphi(t) - \int_{t\wedge\tau_{N+1}} ^{T\wedge\tau_{N+1}} \theta(u)  dY_{N+1}(u).
\end{align}
Since $\varphi$ is mean self-financing, $C^{\varphi}$ is a $\Gx$-martingale which implies that $V^\varphi (t)+ \Theta(t)$  is a $\Gx$-martingale by applying Eq.~\eqref{eq:cost}. Notice that $V^{\varphi}(\tau_{N+1} \wedge T)+\Theta(\tau_{N+1} \wedge T)=\Theta(\tau_{N+1} \wedge T)$. Then it holds that, for $t \in [0, T\wedge\tau_{N+1}]$,
\begin{align*}
V^\varphi (t) + \Theta(t) = \Ex \left[  \Theta(T\wedge\tau_{N+1}) | \G_t \right].
\end{align*}
Using Eq.~\eqref{eq:GKWbis}, we get for $t \in [0, T\wedge\tau_{N+1}]$  that
\begin{align*}
V^\varphi (t) + \Theta(t) = \Ex \left[  \Theta(T\wedge\tau_{N+1})\right] + \int_{0} ^{t} \theta^{GKW}(u)  dY_{N+1}(u) + A(t).
\end{align*}
Plugging the above display into Eq.~(\ref{conti}) and we arrive at
$$C^{\varphi}(T\wedge\tau_{N+1}) - C^{\varphi}(t)= \int_{t\wedge\tau_{N+1}} ^{T\wedge\tau_{N+1}} (\theta^{GKW} (u)  - \theta(u) )dY_{N+1}(u)  + A(T\wedge\tau_{N+1}) - A(t).$$
Analogously, it holds that for $t \in[0, T\wedge\tau_{N+1}]$,
\begin{align*}
C^{\varphi^*}(T\wedge\tau_{N+1}) - C^{\varphi^*}(t)=A(T\wedge\tau_{N+1}) - A(t).
\end{align*}
As a consequence, for $t \in[0, T\wedge\tau_{N+1}]$, and using the strong orthogonality between $Y_{N+1}$ and $A$, we obtain that
\begin{align}\label{eq:Rvarphi}
& R^{\varphi}(t) = R^{\varphi^*}(t) + \Ex \left[  \left (\int_{t\wedge\tau_{N+1}} ^{T\wedge\tau_{N+1}} (\theta^{GKW} (u)  - \theta(u) )dY_{N+1}(u) \right )^2  \bigg| \G_t \right]\nonumber\\
&\qquad\qquad +
\nonumber \Ex \left[ ( A(T\wedge\tau_{N+1}) - A(t)) \left (\int_{t\wedge\tau_{N+1}} ^{T\wedge\tau_{N+1}} [\theta^{GKW} (u)  - \theta(u)]  dY_{N+1}(u) \right) \bigg| \G_t \right]
\\
& \qquad\quad  = R^{\varphi^*}(t) + \Ex \left[  \left (\int_{t\wedge\tau_{N+1}} ^{T\wedge\tau_{N+1}} (\theta^{GKW} (u)  - \theta(u) )dY_{N+1}(u) \right )^2  \bigg| \G_t \right].
\end{align}
Thus it holds that
\begin{align*}
R^{\varphi}(t) \geq R^{\varphi^*}(t),\qquad \forall\ t \in[0, T\wedge\tau_{N+1}],
\end{align*}
for any mean self-financing  and $0$-achieving strategy $\varphi = (\theta, \eta)\in \Psi$.

 We now prove the uniqueness. If there exists a different mean self-financing and $0$-achieving strategy $\varphi = (\theta, \eta)\in \Psi$, which is also risk-minimizing, then by Eq.~\eqref{eq:Rvarphi} we obtain
\begin{align*}
\Ex \left[  \left |\int_{0} ^{T\wedge\tau_{N+1}} (\theta^{GKW} (u)  - \theta(u) ) dY_{N+1}(u) \right |^2  \right] = \Ex \left[ \int_{0} ^{T\wedge\tau_{N+1}} (\theta^{GKW} (u)  - \theta(u))^2 d\left<Y_{N+1},Y_{N+1}\right>  \right] =0
\end{align*}
which implies $\theta^{GKW}(t) = \theta(t)$, for a.e. $t \in [0, T\wedge\tau_{N+1}]$.
Finally Eq.~\eqref{eq:thetaGKW2} follows by Eq.~\eqref{eq:GKWbis} and the orthogonality between $Y_{N+1}$ and $A$.} \hfill$\Box$\\

\noindent{\bf Proof of Theorem~\ref{thm:sol}.}\quad Without loss of generality, we set $L_{N+1}(z)=1$ for all $z\in{\cal S}$. Then in the the default state $z=0^{j_1,\ldots,j_l}$, we may rewrite Eq.~\eqref{eq:F-K-g2z=m} in the following abstract form: on $(t,x)\in[0,T)\times\R_+^{N+1}$,
\begin{align}\label{eq:eqm}
\left(\frac{\partial}{\partial t}+\tilde{\cal A}\right)u(t,x)+h(x)u(t,x)+w(t,x)=0
\end{align}
with terminal condition $u(T,x)=0$ for all $x\in\R_+^{N+1}$. The coefficients are given by
\begin{align*}
h(x)&:=-\sum_{j\notin\{j_1,\ldots,j_l\}} x_j,\nonumber\\
w(t,x)&:=\left\{\sum_{i=1}^{\bar{N}} b_i(1-K_i^{(l+1),N+1})\big[F_{(1,1,1)i}(t,x+w_{N+1},0^{j_1,\ldots,j_l,N+1})-Z_i^{(l+1),N+1}K_i^{(l+1),N+1}\big]\right\}_+\nonumber\\
&\qquad\times x_{N+1}{\bf1}_{j_1,\ldots,j_l\neq N+1}+\sum_{j\notin\{j_1,\ldots,j_l\}}g^{(l+1),j}(t,x+w_j)x_j.
\end{align*}
We also want to apply Theorem 1 of \cite{health2000} to prove existence and uniqueness of classical solutions of PDE~\eqref{eq:eqm} by verifying that the series of conditions [A1], [A2], [A3'] and [A3a']-[A3e'] hold in our case. To this purpose, we first consider bounded domains $D_n:=(\frac{1}{n},n)^{N+1}$, $n\in\N$, with smoothed corners so that they satisfy $\cup_{n=1}^{\infty}D_n=\R_+^{N+1}$. We can then verify that the condition [A3'] holds in the domain of the equation. Using assumptions ({\bf A1}) and ({\bf A2}), the conditions [A1] and [A2] hold. The same assumption also implies that [A3a'] holds. Moreover $\sigma\sigma^{\top}(x)$ is uniformly elliptic on $(t,x)\times\overline{D}_n$, i.e. [A3b'] holds.
Notice that the solution $g^{(l+1),j}(t,x+w_j)$ is bounded and $C^{1,2}$ in $(t,x)$ by the induction hypothesis for $j\notin\{j_1,\ldots,j_l\}$. The function $F_{(1,1,1)i}(t,x)$ is also bounded and $C^{1,2}$ in $(t,x)$ for $i=1,\ldots,\bar{N}$ by Proposition~\ref{lem:sol-Fi2}. Notice that $h(x)$ is linear in $x$. Then the conditions [A3c'] and [A3d'] on the coefficients $h(x)$ and $w(t,x)$,  $(t,x)\in[0,T]\times\overline{D}_n$, are satisfied. It is left to verify [A3e']. For this, it suffices to prove the uniform integrability of the family
\begin{align}\label{eq:ui0}
\left\{\int_t^Tw(s,\tilde{X}^{(t,x)}(s))e^{-\int_t^sh(\tilde{X}^{(t,x)}(u))du}ds;\ (t,x)\in[0,T]\times\R_+^{N+1}\right\}.
\end{align}
Above, the underlying $N+1$-dimensional $\R_+^{N+1}$-valued process $\tilde{X}^{(t,x)}(t)=(\tilde{X}_j^{(t,x)}(t))^{\top}_{j=1,\ldots,N+1}$ satisfies SDE~\eqref{eq:Xtilde}. Consider first the case $N+1\in\{j_1,\ldots,j_l\}$. Because the function $g^{(l+1),j}(t,x)$ is bounded on $[0,T]\times\R_+^{N+1}$ by the induction hypothesis, there exists a constant $C>0$ such that
\begin{align*}
&\Ex\left[\left|\int_t^Tw(s,\tilde{X}^{(t,x)}(s))e^{\int_t^sh(\tilde{X}^{(t,x)}(u))du}ds\right|^2\right]\nonumber\\
&\quad\leq C\Ex\left[\left|\int_t^T\left(\sum_{j\notin\{j_1,\ldots,j_l\}}\tilde{X}_j^{(t,x)}(s)\right)e^{-\int_t^s\sum_{k\notin\{j_1,\ldots,j_l\}} \tilde{X}_k^{(t,x)}(u)du}ds\right|^2\right]\nonumber\\
&\quad=C\Ex\left[\left|\int_t^Te^{-\int_t^s\sum_{k\notin\{j_1,\ldots,j_l\}} \tilde{X}_k^{(t,x)}(u)du}d\left(\int_t^s\sum_{j\notin\{j_1,\ldots,j_l\}}\tilde{X}_j^{(t,x)}(u)du\right)\right|^2\right]\nonumber\\
&\quad\leq C\left\{1+\left|\Ex\left[e^{-\int_t^T\sum_{k\notin\{j_1,\ldots,j_l\}} \tilde{X}_k^{(t,x)}(u)du}\right]\right|^2\right\}\nonumber\\
&\quad \leq C,
\end{align*}
where $C>0$ and independent of $(t,x)$. Next, consider the case $N+1\notin\{j_1,\ldots,j_l\}$. Also notice that the function $F_{(1,1,1)i}(t,x)$ is also bounded and $C^{1,2}$ in $(t,x)$ for $i=1,\ldots,\bar{N}$ by Proposition~\ref{lem:sol-Fi2}. It follows that there exists a constant $C>0$ such that
\begin{align*}
&\Ex\left[\left|\int_t^Tw(s,\tilde{X}^{(t,x)}(s))e^{\int_t^sh(\tilde{X}^{(t,x)}(u))du}ds\right|^2\right]\nonumber\\
&\quad\leq C\Ex\left[\left|\int_t^Te^{-\int_t^s\sum_{k\notin\{j_1,\ldots,j_l\}} \tilde{X}_k^{(t,x)}(u)du}\left(\tilde{X}_{N+1}^{(t,x)}(s)+\sum_{j\notin\{j_1,\ldots,j_l\}}\tilde{X}_j^{(t,x)}(s)\right)ds\right|^2 \right].
\end{align*}
Using the assumption ({\bf A2}) and since $N+1\in\{j_1,\ldots,j_l\}^c$, it holds that $\tilde{X}_{N+1}^{(t,x)}(s)\leq \sum_{k\notin\{j_1,\ldots,j_l\}} \tilde{X}_k^{(t,x)}(s)$, a.s..
This implies that
\begin{align*}
&\Ex\left[\left|\int_t^Tw(s,\tilde{X}^{(t,x)}(s))e^{\int_t^sh(\tilde{X}^{(t,x)}(u))du}ds\right|^2\right]\nonumber\\
&\quad\leq 4C\Ex\left[\left|\int_t^Te^{-\int_t^s\sum_{k\notin\{j_1,\ldots,j_l\}} \tilde{X}_k^{(t,x)}(u)du}d\left(\int_t^s\sum_{j\notin\{j_1,\ldots,j_l\}}\tilde{X}_j^{(t,x)}(u)du\right)\right|^2\right]\nonumber\\
&\quad\leq 4C\left\{1+\left|\Ex\left[e^{-\int_t^T\sum_{k\notin\{j_1,\ldots,j_l\}} \tilde{X}_k^{(t,x)}(u)du}\right]\right|^2\right\}\nonumber\\
&\quad \leq 4C,
\end{align*}
where $C>0$ and independent of $(t,x)$. Thus we have verified the existence of a constant $C>0$, independent of $(t,x)$, such that
\begin{align*}
\sup_{(t,x)\in[0,T]\times\R_+^{N+1}}\Ex\left[\left|\int_t^Tw(s,\tilde{X}^{(t,x)}(s))e^{\int_t^sh(\tilde{X}^{(t,x)}(u))du}ds\right|^2\right]\leq C<+\infty.
\end{align*}
This yields the uniform integrability of the family \eqref{eq:ui0}. It implies that the condition [A3e'] of \cite{health2000} holds. Using Theorem 1 of \cite{health2000}, we conclude that Eq.~\eqref{eq:eqm} admits a unique classical solution $u(t,x)$ on $[0,T]\times\R_+^{N+1}$.

Next, we prove the solution is nonnegative and bounded on $[0,T]\times\R_+^{N+1}$. We first write the Feymann-Kac's representation of the classical solution $u(t,x)$. For all $(t,x)\in[0,T]\times\R_+^{N+1}$,
\begin{align}\label{eq:FKsolutionm}
u(t,x)&=\Ex\Bigg[\int_t^Te^{-\int_t^s\sum_{k\notin\{j_1,\ldots,j_l\}} \tilde{X}_k^{(t,x)}(u)du}\Bigg(\sum_{j\notin\{j_1,\ldots,j_l\}}\tilde{X}_j^{(t,x)}(s)g^{(l+1),j}(t,\tilde{X}_j^{(t,x)}(s)+w_j)\nonumber\\
&\qquad+\left\{\sum_{i=1}^{\bar{N}} b_i(1-K_i^{(l+1),N+1})\big[F_{(1,1,1)i}(t,x+w_{N+1},0^{j_1,\ldots,j_l,N+1})-Z_i^{(l+1),N+1}K_i^{(l+1),N+1}\big]\right\}_+\nonumber\\
&\qquad\times \tilde{X}_{N+1}^{(t,x)}(s){\bf1}_{j_1,\ldots,j_l\neq N+1}\Bigg)ds \Bigg].
\end{align}
 Moreover, if $N+1\in\{j_1,\ldots,j_l\}$, then Eq.~\eqref{eq:FKsolutionm} reduces to
\begin{align*}
u(t,x)&=\Ex\left[\int_t^Te^{-\int_t^s\sum_{k\notin\{j_1,\ldots,j_l\}} \tilde{X}_k^{(t,x)}(u)du}\left(\sum_{j\notin\{j_1,\ldots,j_l\}}\tilde{X}_j^{(t,x)}(s)g^{(l+1),j}(t,\tilde{X}_j^{(t,x)}(s)+w_j)\right)ds\right].
\end{align*}
Since the nonnegative function $g^{(l+1),j}(t,x)$ is bounded on $[0,T]\times\R_+^{N+1}$ by the inductive hypothesis, there exists a constant $C>0$ such that
\begin{align*}
0\leq u(t,x)&\leq C\sum_{j\notin\{j_1,\ldots,j_l\}}\Ex\left[\int_t^T\tilde{X}_j^{(t,x)}(s)e^{-\int_t^s\sum_{k\notin\{j_1,\ldots,j_l\}} \tilde{X}_k^{(t,x)}(u)du}ds\right]\nonumber\\
&=C\Ex\left[\int_t^Te^{-\int_t^s\sum_{k\notin\{j_1,\ldots,j_l\}} \tilde{X}_k^{(t,x)}(u)du}d\left(\int_t^s\sum_{j\notin\{j_1,\ldots,j_l\}}\tilde{X}_j^{(t,x)}(u)du\right)\right]\nonumber\\
&=C\left\{1-\Ex\left[e^{-\int_t^T\sum_{k\notin\{j_1,\ldots,j_l\}} \tilde{X}_k^{(t,x)}(u)du}\right]\right\}.
\end{align*}
Obviously, the above inequality yields the existence of a constant $C>0$ such that $0\leq u(t,x)\leq C$ for all $(t,x)\in[0,T]\times\R_+^{N+1}$.
Next, consider the case $N+1\notin\{j_1,\ldots,j_l\}$. It follows from \eqref{eq:FKsolutionm} that
\begin{align*}
u(t,x)&=\Ex\Bigg[\int_t^Te^{-\int_t^s\sum_{k\notin\{j_1,\ldots,j_l\}} \tilde{X}_k^{(t,x)}(u)du}\Bigg(\sum_{j\notin\{j_1,\ldots,j_l\}}\tilde{X}_j^{(t,x)}(s)g^{(l+1),j}(t,\tilde{X}_j^{(t,x)}(s)+w_j)\nonumber\\
&\qquad+\left\{\sum_{i=1}^{\bar{N}} b_i(1-K_i^{(l+1),N+1})\big[F_{(1,1,1)i}(t,x+w_{N+1},0^{j_1,\ldots,j_l,N+1})-Z_i^{(l+1),N+1}K_i^{(l+1),N+1}\big]\right\}_+\nonumber\\
&\qquad\times \tilde{X}_{N+1}^{(t,x)}(s)\Bigg)ds \Bigg].
\end{align*}
Using the assumption ({\bf A2}), there exists a constant $C>0$ such that
\begin{align*}
0\leq u(t,x)&\leq C\Ex\left[\int_t^Te^{-\int_t^s\sum_{k\notin\{j_1,\ldots,j_l\}} \tilde{X}_k^{(t,x)}(u)du}\left(\sum_{j\notin\{j_1,\ldots,j_l\}}\tilde{X}_j^{(t,x)}(s)+\tilde{X}_{N+1}^{(t,x)}(s)\right)ds \right].
\end{align*}
Since $N+1\in\{j_1,\ldots,j_l\}^c$, it follows from the assumption ({\bf A2}) that $\tilde{X}_{N+1}^{(t,x)}(s)\leq \sum_{k\notin\{j_1,\ldots,j_l\}} \tilde{X}_k^{(t,x)}(s)$, a.s.. This implies that
\begin{align*}
0\leq u(t,x)&\leq 2C\Ex\left[\int_t^Te^{-\int_t^s\sum_{k\notin\{j_1,\ldots,j_l\}} \tilde{X}_k^{(t,x)}(u)du}d\left(\int_t^s\sum_{j\notin\{j_1,\ldots,j_l\}}\tilde{X}_j^{(t,x)}(u)du\right)\right]\nonumber\\
&=2C\left\{1-\Ex\left[e^{-\int_t^T\sum_{k\notin\{j_1,\ldots,j_l\}} \tilde{X}_k^{(t,x)}(u)du}\right]\right\}.
\end{align*}
Obviously, the above inequality gives a constant $C>0$ such that $0\leq u(t,x)\leq C$ for all $(t,x)\in[0,T]\times\R_+^{N+1}$. This completes the proof of the theorem. \hfill$\Box$

\end{document}